\documentclass[english, aps,prd,twocolumn,twoside,superscriptaddress,floatfix, nofootinbib]{revtex4-2}

\usepackage[utf8]{inputenc}
\usepackage{amsmath}
\usepackage{geometry}
\usepackage{xcolor}
\usepackage{graphicx}
\usepackage{float}
\usepackage{lipsum, babel}
\usepackage{fontawesome5}

\newcommand{\github}[1]{%
   \href{#1}{\faGithubSquare}%
}

\usepackage{comment}

\usepackage{color}
\usepackage{hyperref}
\usepackage{ifthen,amsmath,amssymb, bm}

\newcommand{\Beam}[0]{\ensuremath{\mathcal B}}

\newcommand{\n} {\hat{{n}}}

\newcommand{\beq} {\begin{equation}}
\newcommand{\eeq} {\end{equation}}
\newcommand{\bal} {\begin{aligned}}
\newcommand{\eal} {\end{aligned}}

\newcommand{\Geneve}{Universit\'e de Gen\`eve, D\'epartement de Physique Th\'eorique et CAP, 24 Quai Ansermet, CH-1211 Gen\`eve 4, Switzerland}

\begin{document}

\title{Optimal joint reconstruction from CMB observations: application to cosmic birefringence, patchy reionization and CMB lensing}

\author{Omar Darwish}
\email{od261@cantab.ac.uk}
\affiliation{\Geneve}

\begin{abstract}
Line-of-sight distortions of the cosmic microwave background (CMB), including gravitational lensing, cosmic birefringence, and patchy screening, encode crucial cosmological information. While quadratic estimators (QE) have been excellent tools for extracting these signals, they become suboptimal for current- and next-generation CMB surveys, failing to maximize signal-to-noise and suffering from bias contamination that standard bias-hardening techniques cannot mitigate. We present a joint maximum a posteriori framework that simultaneously reconstructs multiple distortion fields while explicitly accounting for their mutual contamination. For cosmic birefringence searches, our method achieves up to 2.5 improvement in reconstruction noise compared to the QE, while significantly reducing CMB lensing-induced biases up to a factor of 5. These gains in lensing biases manifest not only in deep polarization surveys like CMB-S4 and SPT-3G, but also in higher-noise experiments like Simons Observatory, where our method reduces lensing-induced biases by a factor of two thanks to the power of delensing. Our code provides a first step to robust analyses of CMB secondary anisotropies, their cross-correlations with large-scale structure, and ultimately enabling more sensitive searches for primordial $B$-modes \href{https://github.com/Saladino93/jointmap}{\color{blue}\faGithub}.
\end{abstract}

\maketitle

\section{Introduction}

Next-generation CMB surveys will precisely map cosmic microwave background polarization, with Simons Observatory (SO) \cite{thesimonsobservatorycollaborationSimonsObservatoryScience2019}, CMB Stage 4 (CMB-S4) \cite{abazajianCMBS4ScienceBook2016a}, BICEP3/Keck Array \cite{Grayson_2016}, and SPT-3G \cite{bensonSPT3GNextGenerationCosmic2014} aiming to detect primordial $B$-mode polarization from inflationary gravitational waves \cite{kamionkowskiQuestModesInflationary2016}.

However, post-recombination physics can also generate B-modes, even in the absence of primordial ones, within and beyond the Standard Model of physics \cite{changSnowmass2021CosmicFrontier2022b}. This includes early-time cosmic birefringence rotating the CMB linear polarization \cite{marshAxionCosmology2016, kamionkowskiHowDeRotateCosmic2009, gluscevicDeRotationCosmicMicrowave2009, namikawaTestingParityviolatingPhysics2019}, patchy reionization suppressing CMB anisotropies \cite{gruzinovSecondaryCMBAnisotropies1998, huReionizationRevisitedSecondary2000a, santosSmallscaleCMBTemperature2003, dvorkinReconstructingPatchyReionization2009}, and large-scale structure gravitationally lensing the CMB \cite{lewisWeakGravitationalLensing2006}. Each distortion encodes valuable cosmological information, making their accurate measurement crucial.

\begin{figure}[htbp]
    \centering
    \includegraphics[width=\linewidth]{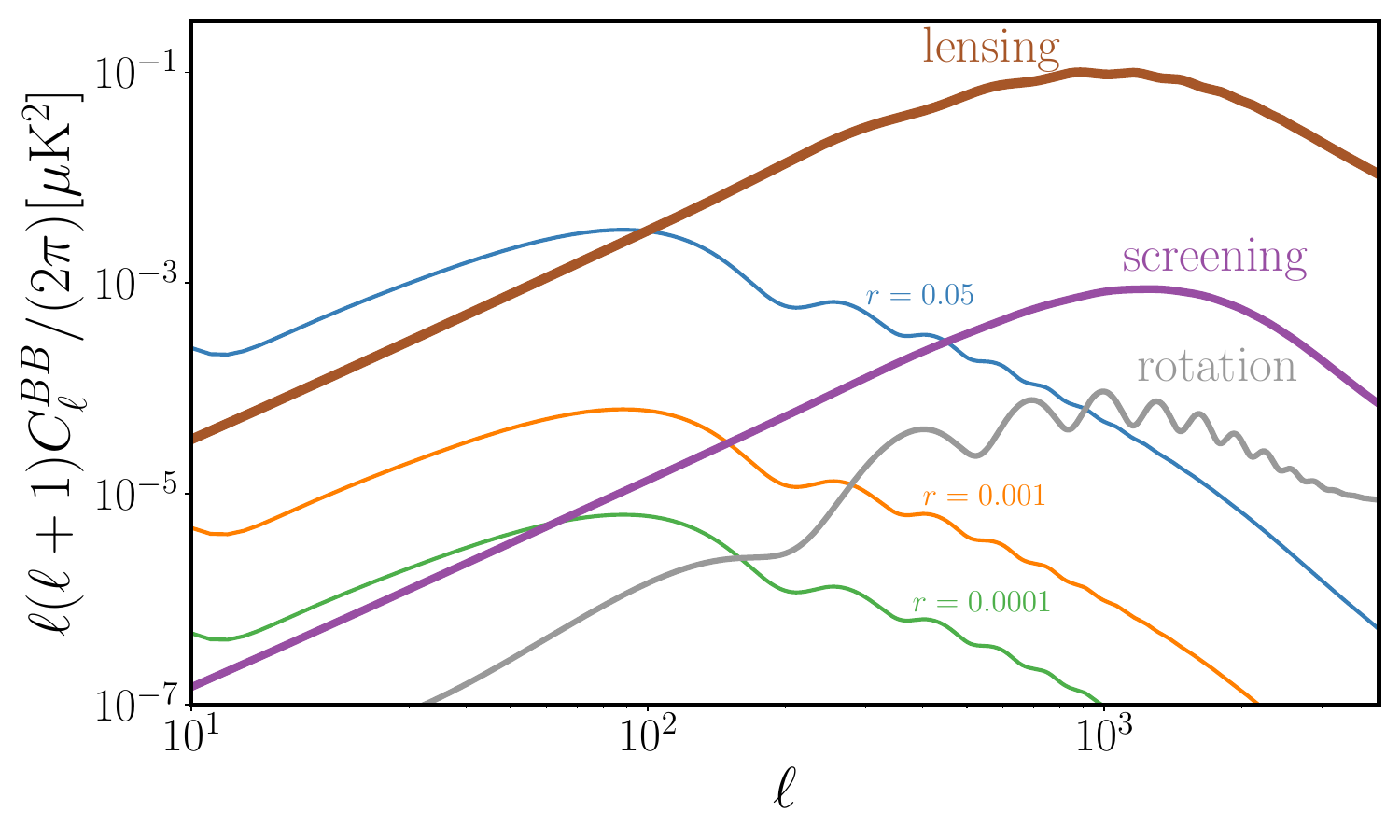}
    \caption{Comparing $B$-modes spectra generated by different physical processes. In the absence of primordial $B$-modes we show late-time lensing, screening, and rotation $B$-modes, in brown, purple and grey, respectively. We clearly see that the lensing-induced B-modes dominate the reconstruction noise of Equation \ref{eq:recnoisesqueezed}. For reference, we show primordial tensor $B$-modes with different levels of the tensor-to-scalar ratio $r$ (at a pivot scale $k=0.05\ \mathrm{Mpc}^{-1}$). We use \texttt{CAMB} to generate the tensor, lensed and scalar unlensed spectra \cite{Lewis:1999bs}.\footnote{\url{camb.readthedocs.io}} We use \texttt{lenspyx} with expressions in \cite{dvorkinReconstructingPatchyReionization2009, Cai_2022} to calculate the screening only and rotation CMB induced $B$-modes.\footnote{\url{https://github.com/carronj/lenspyx}} }
    \label{fig:bbspectra}
\end{figure}

While quadratic estimators (QE) have been the standard tool for analyzing these effects \cite{huMassReconstructionCMB2002, gluscevicDeRotationCosmicMicrowave2009, kamionkowskiHowDeRotateCosmic2009, dvorkinReconstructingPatchyReionization2009, gluscevicPatchyScreeningCosmic2013, carronPositionspaceCurvedskyAnisotropy2019, maniyarQuadraticEstimatorsCMB2021}, they face fundamental limitations. For low-noise experiments, QE estimators hit a noise floor set by lensed, rotated and modulated spectra. The polarization-only $EB$ estimator noise in the squeezed limit demonstrates this ceiling in the reconstruction noise from Gaussian chance fluctuations \cite{challinorExploringCosmicOrigins2018, carronSphericalBispectrumExpansion2024}:
\begin{equation}
    N_{0,L}^{\mathrm{QE}} \sim \left(\sum_{l} \frac{2l+1}{4\pi}\frac{(C_l^{EE})^2}{C^{EE}_{l,\mathrm{tot}} C^{BB}_{l, \mathrm{tot}}} \right)^{-1}\ ,\label{eq:recnoisesqueezed}
\end{equation}
where the total distortion-induced $B$-modes in $C^{BB}_{l, \mathrm{tot}}$ fundamentally limit the achievable noise \cite{hirataReconstructionLensingCosmic2003, lewisWeakGravitationalLensing2006}, and $C_l^{EE}$, $C^{EE}_{l,\mathrm{tot}}$ are the $E$ mode distorted and total spectra, respectively. Figure \ref{fig:bbspectra} shows that CMB lensing is by far the largest source of late-time induced B-modes that needs to be mitigated to improve the reconstruction noise.

Moreover, different distortions contaminate each other's measurements through biases in the QE power spectra \cite{suImprovedForecastPatchy2011, namikawaTestingParityviolatingPhysics2019, caiImpactAnisotropicBirefringence2023b}. While bias-hardening techniques can address some field-level biases \cite{suImprovedForecastPatchy2011, namikawaBiashardenedCMBLensing2013}, they fail for biases due to secondary CMB four-point contractions, known as $N_1$ biases \cite{Kesden_2003, namikawaTestingParityviolatingPhysics2019, caiImpactAnisotropicBirefringence2023b}. For example, the cosmic birefringence power spectrum picks up a CMB lensing induced bias. In the squeezed limit, this $N_1$ bias, for the QE $EB$ estimator, can be approximated as
\cite{carronDetectingRotationLensing2025c}
\begin{equation}
    N_{1,L}^{\mathrm{QE}} \sim \sum_{L'} C_{L'}^{\phi\phi} G(L,L',C^{EE},C^{EE}_{\mathrm{tot}},C^{BB}_{\mathrm{tot}})\ ,
\end{equation}
summing only over the large scales $\sum_{L'}$ $C_{L'}^{\phi\phi}$ of the CMB lensing power spectrum, and $G$ is some function that captures the coupling between small-scales $E$ and $B$ due to the presence of lensing. If we can somehow remove lensing from CMB maps (delensing), we might increase the SNR of CMB anisotropies estimators while mitigating lensing induced biases.

Likelihood-based methods offer a solution by jointly maximizing SNR and mitigating biases. This has been demonstrated in maximum a posteriori (MAP) CMB lensing reconstruction \cite{hirataAnalyzingWeakLensing2003, hirataReconstructionLensingCosmic2003, Carron_2017, milleaBayesianDelensingDelight2020, legrandLensingPowerSpectrum2022, legrandRobustEfficientCMB2023, darwish2024nongaussiandeflectionsiterativeoptimal}, that recently achieved state-of-the-art measurements with SPT-3G polarization data \cite{milleaOptimalCosmicMicrowave2021, geCosmologyCMBLensing2024}, at a competitive level with measurements using a QE combination of temperature and polarization data from the Planck and the Atacama Cosmology Telescope collaborations \cite{planckcollaborationPlanck2018Results2020, quAtacamaCosmologyTelescope2023}. While future applications to CMB-S4 \cite{Belkner:2023duz} will likely benefit the most from likelihood-based reconstruction, we show here that even surveys not dominated by polarization data, like SO, can gain value from likelihood methods (or just simple delensing). In the short-term, nominal SPT-3G, with observations up to 2026 well before CMB-S4, will be the one to substantially benefit from these methods \cite{prabhu2024testingmathbflambdacdmcosmologicalmodel}.

We present a general MAP framework for joint reconstruction of multiple CMB distortion fields, building upon previous work \cite{Carron_2017}, extending the \texttt{delensalot} code \cite{Belkner:2023duz}.\footnote{\url{https://github.com/Saladino93/delensalotlight}, fork of \url{https://github.com/NextGenCMB/delensalot}, though in the future \texttt{delensalot} will receive big upgrades.} Our method simultaneously accounts for mutual contamination while maximizing SNR, enabling robust analyses. As a side product, the method allows to reduce residual B-modes by derotating, de-screening, and delensing the observed CMB (a term that we generally label “de-operating”).

Previous works have explored similar directions using the QE \cite{yadavPrimordialBmodeDiagnostics2009a, fengSearchingPatchyReionization2018}, maximum a posteriori methods \cite{bianchiniInferenceGravitationalLensing2023}, and machine learning methods \cite{guzmanReconstructingPatchyReionization2021, guzmanReconstructingCosmicPolarization2022}. While machine learning methods are still under development, the more mature QE approach currently deals with simulation based methods to mitigate independently these biases, as recently done by the BICEP/Keck collaboration\cite{collaborationBICEPKeckXVII2023}. Our work goes into the direction of unifying CMB distortion analyses in one single framework.\footnote{The most similar work to ours is \cite{bianchiniInferenceGravitationalLensing2023}, where they jointly sample the unlensed CMB, the CMB lensing potential, and patchy tau \cite{milleaMUSEMarginalUnbiased2021a}. While conceptually the method presented here and there are similar, and can both saturate the information from data, the differences are practical. In its current form the former method is more robust to mis-modeling, as it can account for a mean-field and RD-$N_0$ subtraction, though these in principle can be included in \cite{milleaMUSEMarginalUnbiased2021a}. For an extensive discussion see the recent work of \cite{carron2025likelihoodbasedanalysisgravitationallydelensed}.} 

The paper is organized as follows: Section \ref{sec:distortingphysics} reviews CMB distortion physics and quadratic estimators. Section \ref{sec:firstexperiments} demonstrates QE limitations in cosmic birefringence reconstruction. Sections \ref{sec:jointmaprec} and \ref{sec:results} present our joint MAP framework and its application. We conclude in Section \ref{sec:conclusions}.

\section{CMB Distortion physics extracted with quadratic estimators \label{sec:distortingphysics}}

While the primordial CMB exhibits statistical isotropy, various physical processes introduce anisotropies by modifying its covariance matrix \cite{Hanson_2009}. These distortions can be reconstructed using quadratic estimators (QE) \cite{huMassReconstructionCMB2002}, whose derivation we sketch here following \cite{hirataAnalyzingWeakLensing2003, Hanson_2009, hansonWeakLensingCMB2010, carronPositionspaceCurvedskyAnisotropy2019}.

Consider a distorting field $\xi_{j}$ acting through an operator $O_j$ on the CMB. The observed temperature and polarization components $X^{\mathrm{dat}} = (T^{\mathrm{dat}}, Q^{\mathrm{dat}}, U^{\mathrm{dat}})$ are modeled as:
\begin{equation}
X^{\mathrm{dat}}(\hat{n}) = \Beam O_j X + n,
\end{equation}
where $X$ represents primordial CMB anisotropies, $\Beam$ the instrumental beam, and $n$ uncorrelated instrumental noise. The CMB log-likelihood for a fixed distortion $\xi_j$ is (assuming Gaussian noise) \cite{hansonWeakLensingCMB2010}:
\begin{multline}
L(X^{\mathrm{dat}}|\xi_j) \equiv -\frac{1}{2} X^{\mathrm{dat},\dagger}\mathrm{Cov}_j^{-1}X^{\mathrm{dat}} + \\ - \frac{1}{2}\ln \det \mathrm{Cov}_j + \mathrm{const}\ ,\label{eq:loglike}
\end{multline}
where the covariance matrix is
\begin{equation}
   \mathrm{Cov}_j =  \Beam O_j C^{u}  O_j^{\dagger} \Beam^{\dagger} + N,
\end{equation}
with $C^{u},N$ the primordial CMB and noise covariances, respectively.
Let's now maximize \ref{eq:loglike} with respect to $\xi_j$ (note that $\mathrm{Cov}_j \mathrm{Cov}_j^{-1} = \mathbf{I}$):

\begin{equation}
\frac{\delta L}{\delta \xi_j(\hat{y})} = \frac{1}{2} \bar{X_j}^{\dagger} \frac{\delta \mathrm{Cov}_j}{ \delta \xi_j(\hat{y})} \bar{X_j}\ - \frac{1}{2}\frac{\delta \ln \det \mathrm{Cov}_j}{\delta \xi_j(\hat{y})} =0, \label{eq:derloglike}
\end{equation}
where $\bar{X_j} = \mathrm{Cov}_j^{-1} X^{\mathrm{dat}}$ is an inverse-variance filtered CMB map. In this expression, the first term is a quadratic part $g_{\mathrm{QD}}$, and $\frac{\delta \ln \det \mathrm{Cov}_j}{\delta \xi_j}=\frac{1}{2}\mathrm{Tr}[\mathrm{Cov}_j^{-1}\frac{\delta \mathrm{Cov}_j}{ \delta \xi_j}] =: \bar{\xi}_j $ gives rise to a QE mean-field term $\bar{\xi}_j$ accounting for other sources of CMB anisotropy, such as  masking and the source of anisotropy itself \cite{Benoit_L_vy_2013, Carron_2017}. Usually it is estimated by averaging the QE over simulations. As we will work in a full-sky setting and set $\xi_j=0$ in the covariance matrix, we can ignore the mean-field term.

Consider expanding the log-likelihood gradient around $\xi_j=0$, assuming small anisotropies. This expansion allows us to solve for the maximum likelihood estimate of a single distorting field:
\begin{equation}
    0 = \frac{\delta L}{\delta \xi_j(\hat{y})} \approx  \frac{\delta L}{\delta \xi_j(\hat{y})}\Big\vert_{\xi_j=0}+ \frac{\delta^2 L}{\delta^2 \xi_j}\Big\vert_{\xi_j=0} \xi_j\ .\label{eq:qeliketozero}
\end{equation}
This leads to
\begin{equation}
    \xi_j \approx -\left(\frac{\delta^2 L}{\delta^2\xi_j}\Big\vert_{\xi_j=0}\right)^{-1}\left(\frac{1}{2} \bar{X_j}^{\dagger} \frac{\delta \mathrm{Cov}_j}{ \delta \xi_j(\hat{y})}\Big\vert_{\xi_j=0} \bar{X_j}-\bar{\xi}_j\Big\vert_{\xi_j=0}\right)\ .
\end{equation}
The second derivative in the parentheses is approximated with the expectation value over data, the Fisher matrix (note that by definition $\int_{X} e^{L}=1$):
\begin{multline}
    -\Big\langle \frac{\delta^2 L}{\delta^2 \xi_j}\Big\vert_{\xi_j=0} \Big\rangle = \\ = \frac{1}{2}\mathrm{Tr}[\frac{\delta \mathrm{Cov}_j}{\delta \xi_j}\mathrm{Cov}_j^{-1}\frac{\delta \mathrm{Cov}_j}{\delta \xi_j}\mathrm{Cov}_j^{-1}]\Big\vert_{\xi_j=0} \equiv R
\end{multline}
where we have defined the response $R$.

Hence, our quadratic estimator is
\begin{equation}
\widehat{\xi_j}^{\mathrm{QE}}=R^{-1}\left(\frac{1}{2} \bar{X_j}^{\dagger} \frac{\delta \mathrm{Cov}_j}{ \delta \xi_j(\hat{y})}\Big\vert_{\xi_j=0} \bar{X_j}-\bar{\xi}_j\Big\vert_{\xi_j=0}\right) \label{eq:likeqe}\ ,
\end{equation}
where we evaluate the response, the inverse covariance matrix, and the derivative of the covariance with respect to $\xi_j$ at zero distortion. 
Taking the expectation value over primordial CMB of this QE, leads to
\begin{multline}
    \langle \widehat{\xi_j}^{\mathrm{QE}} \rangle_{\mathrm{CMB}} = \\ = R^{-1} \left(\frac{1}{2}\mathrm{Tr}[\mathrm{Cov}_j^{-1}  \frac{\delta \mathrm{Cov}_j}{ \delta \xi_j(\hat{y})} \mathrm{Cov}_j^{-1} ]\Big\vert_{\xi_j=0}\mathrm{Cov}_j]-\bar{\xi}_j\Big\vert_{\xi_j=0} \right) \\ 
    \approx R^{-1} \frac{1}{2}\mathrm{Tr}[\mathrm{Cov}_j^{-1}  \frac{\delta \mathrm{Cov}_j}{ \delta \xi_j(\hat{y})} \mathrm{Cov}_j^{-1} \frac{\delta \mathrm{Cov}_j}{\delta \xi_j}]\Big\vert_{\xi_j=0}\xi_j =  \xi_j\ ,
\end{multline}
where we approximated the covariance matrix as $\mathrm{Cov}_j \approx \mathrm{Cov}_j\Big\vert_{\xi_j=0}+\frac{\delta \mathrm{Cov}_j}{\delta \xi_j}\Big\vert_{\xi_j=0}\xi_j$ for some small distortion. Therefore, for the QE case here the response $R$ acts as a normalization and a minimum variance reconstruction noise ($\langle\widehat{\xi_j}^{\mathrm{QE}}\widehat{\xi_j}^{\mathrm{QE}}\rangle-\langle\widehat{\xi_j}^{\mathrm{QE}}\rangle\langle\widehat{\xi_j}^{\mathrm{QE}}\rangle$) given by the Fisher matrix in the no distortion case; this QE is unbiased and optimal in this sense. From now on we will ignore the mean-field term.

To compare our QE reconstruction with theoretical predictions, we need to understand the power spectrum of the estimator in the presence of $N$ distortions $\xi_i,\ i \in \{1, ..., N\}$, including $\xi_j$.\footnote{In principle, one can use other statistics, but we will focus on the power spectrum here.} The power spectrum of the QE estimator decomposes into signal and noise terms:

\begin{equation}
\begin{aligned}
C_L^{\widehat{\xi_j}^{\mathrm{QE}}\widehat{\xi_j}^{\mathrm{QE}}} &= C_L^{\xi_j\xi_j}+N_L^{0}+N_L^{1,\widehat{\xi_j}\xi_j}+ \\
&\phantom{=} +\sum_i\Big(C_L^{\widehat{\xi_j}\xi_i} +N_L^{1,\widehat{\xi_j}\xi_i}\Big)  +... \label{eq:totalps}\ .
\end{aligned}
\end{equation}
Here, $C_L^{\xi_j\xi_j}$ is the target signal we aim to measure, $N_L^{0}=R^{-1}$ represents the Gaussian reconstruction noise \cite{huMassReconstructionCMB2002}, and $N_L^{1,\widehat{\xi_j}\xi_j}$ captures the secondary correlations involving the distortion field itself, of order $\mathcal{O}(\xi_j^2)$ \cite{Kesden_2003}.\footnote{$N_0$ has a zero power of the power spectrum $C_L^{\xi_j\xi_j}$. $N_1$ has one power. In general, in the literature $N_n$ has $n$ powers of $C_L^{\xi_j\xi_j}$, where $n$ is a real number.} 

The remaining terms represent contamination from other fields: $C_L^{\widehat{\xi_j}\xi_i}$ describes field-level leakage, while $N_L^{1,\widehat{\xi_j}\xi_i}$ represents higher-order contamination of order $\mathcal{O}(\xi_i^2)$, similar to $N_L^{1,\widehat{\xi_j}\xi_j}$.

To extract the physical power spectrum $C_L^{\xi_j\xi_j}$, modern CMB analysis pipelines must account for these various biases (e.g. \citep{planckcollaborationPlanck2018Results2020}). The Gaussian noise $N_0$ and the higher-order bias $N_L^{1,\widehat{\xi_j}\xi_j}$ are typically subtracted from the total power spectrum using a combination of data and simulations (e.g. \cite{planckcollaborationPlanck2018Results2020}). While the leakage term $C_L^{\widehat{\xi_j}\xi_i}$ could be handled similarly, a more robust approach is to employ bias-hardening techniques, which we describe in the next subsection \cite{namikawaBiashardenedCMBLensing2013, namikawaBiashardenedCMBLensing2014}.

\subsection*{QE Bias-hardening\label{sec:bh}}

Bias-hardening builds estimators robust to contamination from multiple distorting fields. They are widely used in the context of CMB lensing, to deal with masks \cite{namikawaBiashardenedCMBLensing2013} and extra-galactic foreground contamination (e.g. \cite{Osborne_2014, Sailer_2020}). We summarize briefly how they work.

In the presence of multiple distorting fields, the estimator \ref{eq:likeqe} will have an expected value \cite{namikawaBiashardenedCMBLensing2013}:

\begin{equation}
    \langle \widehat{\xi_j}^{\mathrm{QE}} \rangle_{\mathrm{CMB}} = \xi_j + R^{-1}\sum_{i\neq j} R^{\widehat{\xi_j}\xi_i}\xi_i, \label{eq:qeaverage}
\end{equation}
where
\begin{equation}
    R^{\widehat{\xi_j}\xi_i} = \frac{\delta^2 L}{\delta \xi_i\delta \xi_j}\Big\vert_{(\xi_j,\xi_i)=(0,0)}\ 
\end{equation}
represents the response of the estimator to the field $\xi_i$ (note that by definition $R^{\widehat{\xi_j}\xi_j}\equiv R$).

To obtain an unbiased estimator for $\xi_j$ we evaluate Equation \ref{eq:qeliketozero} at multiple fields
\begin{equation}
    \frac{\delta L}{\delta \xi_j}\Big\vert_{\xi=\vec{0}}=\sum_{i=1}^{N}R^{\widehat{\xi_j}\xi_i}\xi_i
\end{equation}
where $\xi=(\xi_1, ...,\xi_N)$.\footnote{We use the notation $\xi$ instead of $\vec{\xi}$ to declutter our equations.} This leads to a system of equations ($\forall\ j \in \{1, ..,N\}$):
\begin{equation}
    R^{\widehat{\xi_j}\xi_j} \widehat{\xi_j}^{\mathrm{QE}}  =  \sum_{i=1}^{N}R^{\widehat{\xi_j}\xi_i}\xi_i\ .
\end{equation}
By working out this system for each distortion estimator $\widehat{\xi_j}^{\mathrm{BH}}$ we get bias-hardened estimators that respect (e.g. \cite{namikawaBiashardenedCMBLensing2013, namikawaConstrainingReionizationFirst2021, darwish2024nongaussiandeflectionsiterativeoptimal})
\begin{equation}
    \langle \widehat{\xi_j}^{\mathrm{BH}} \rangle_{\mathrm{CMB}} = \xi_j\ .
\end{equation}
Hence, these estimators ensure that $C_L^{\widehat{\xi_j}^\mathrm{BH}\xi_i}=0$. On the other hand, a typical $N_1$ contribution is of order $\mathcal{O}(\xi_i^2)$. Writing a bias-hardened estimator as $\widehat{\xi_j}^{\mathrm{BH}} \sim \widehat{\xi}_j^{\mathrm{QE}}-\lambda_{ji}\widehat{\xi_i}^{\mathrm{QE}}$, with $\lambda_{ji}$ some coefficient to ensure the above equation in the case of two distorting fields $\xi_i,\xi_j$, this is not enough to remove beyond $\mathcal{O}(\xi_i)$ biases. The $N_1$ bias contains additional distortions induced CMB correlations not extracted by quadratic estimators; hence the QEs used to build the bias-hardened estimator are insensitive to these extra-correlations at the field level, implying the same for the bias-hardened estimator.

\subsection{Physical Effects and Their Estimators}

We now present three key distortion effects and their QE reconstructions, focusing on polarization data which dominates the signal-to-noise in CMB low-noise regimes \cite{huMassReconstructionCMB2002, hirataReconstructionLensingCosmic2003, bianchiniInferenceGravitationalLensing2023}.

\subsubsection{Amplitude modulation due to reionization}

Inhomogeneous reionization in the early universe leads to a CMB optical depth $\tau(\hat{n})$ varying across the sky \citep{dvorkinReconstructingPatchyReionization2009, gluscevicPatchyScreeningCosmic2013}. Understanding this process of patchy screening is key to gaining insights into the complex process of reionization.\footnote{During the reionization era there are also extra polarization generated on large-scales \cite{dvorkinReconstructingPatchyReionization2009}, and additional temperature CMB anisotropies (from peculiar motion of ionized gas, kSZ effect) \cite{sunyaevObservationsRelicRadiation1972,sunyaevMicrowaveBackgroundRadiation1980}. But these are scattering contributions that we will ignore as we are interested in distortion of the CMB along the line-of-sight.} Along the line of sight, the primordial CMB anisotropies are suppressed by $e^{-\tau(\hat{n})}$, due to the screening of photons, causing a modulation in the observed CMB anisotropies \cite{gruzinovSecondaryCMBAnisotropies1998, schuttNewTemperatureInversion2024}:

\begin{equation}
    X_{\tau}(\hat{n}) = e^{-\tau(\hat{n})} X(\hat{n}) ,\ X\in \{T, Q, U\}\ ,
\end{equation}
transforming screened $E$-modes into new $B$-mode CMB polarization correlated with temperature and $E$-modes \cite{dvorkinReconstructingPatchyReionization2009, gluscevicPatchyScreeningCosmic2013}. 

To get its corresponding QE from \ref{eq:likeqe}, we define an operator $O_{\mathrm{patchy}}(\tau) = e^{-\tau}$, and a covariance $\mathrm{Cov}\equiv O_{\mathrm{patchy}}^{\dagger}C^{u}O_{\mathrm{patchy}}+N$, where $C^{u},N$ are the unlensed CMB and noise covariances, respectively. The quadratic part in Equation \ref{eq:derloglike} gives

\begin{equation}
    g_{\mathrm{QD}}(\hat{n}) = -\bar{X}   \Big(e^{-\tau} X^{\rm{WF}}\Big)(\hat{n})\ , 
\end{equation}
where $X^{\mathrm{WF}}=C^u e^{-\tau} \bar{X}$ is a Wiener-filtered CMB map.

In the limit of no anisotropies except for a small modulation $\delta \tau$, we get an estimator for the optical depth fluctuations given by squared filtered CMB maps

\begin{equation}
    \widehat{\delta \tau}^{\mathrm{QE}} =-R^{-1}\bar{X} \cdot X^{\rm{WF}}\Big\vert_{\tau=0}\ .
\end{equation}




\subsubsection{Cosmic rotation from parity violating physics}

Early universe axionlike particles (ALP) interacting with photons (e.g. \cite{kamionkowskiHowDeRotateCosmic2009, gluscevicDeRotationCosmicMicrowave2009, namikawaTestingParityviolatingPhysics2019}), or primordial magnetic fields through Faraday rotation (e.g. \cite{kosowskyFaradayRotationCosmic2005,pogosianPrimordialMagnetismCMB2012, yadavProbingPrimordialMagnetism2013a, pogosianSearchingPrimordialMagnetic2018a,paolettiLiteBIRDScienceGoals2024}) can induce a rotation of the CMB polarization field (see \cite{namikawaTestingParityviolatingPhysics2019} for a review). Hence, the implications of a detection of cosmic rotation are relevant, as they could imply physics beyond the current Standard Model.

In terms of the Stoke parameters $Q,U$ the CMB polarization $P(\hat{n})=Q(\hat{n})+iU(\hat{n})$ is rotated by an angle $\alpha(\hat{n})$ determined by theory \cite{caldwellCrossCorrelationCosmologicalBirefringence2011}, in such a way that we observe $P(\hat{n})e^{2i\alpha(\hat{n})}$. Using a real space notation, the rotated observed Stokes parameters are \citep{kamionkowskiHowDeRotateCosmic2009}:

\begin{multline}
    \begin{bmatrix}
        T_R(\n) \\
        Q_R(\n) \\
        U_R(\n) 
    \end{bmatrix} = O_{\mathrm{rot}}(\alpha) \begin{bmatrix}
        T(\n) \\
        Q(\n) \\
        U(\n)  
    \end{bmatrix} = \\ = \begin{bmatrix}
        0 & 0 & 0 \\
        0& \cos(2\alpha(\n)) & -\sin(2\alpha(\n)) \\
        0 &  \sin(2\alpha(\n)) & \cos(2\alpha(\n)) 
    \end{bmatrix} \begin{bmatrix}
        T(\n) \\
        Q(\n) \\
        U(\n)   
    \end{bmatrix}\ .
\end{multline}

We can now use Equation \ref{eq:likeqe} to derive a QE estimator for this angle, given CMB data. The relevant covariance in this case is $\mathrm{Cov}=O_{\mathrm{rot}}^{\dagger}C^{\rm{unl}}O_{\mathrm{rot}}+N$. 
In the limit of no anisotropies except for a small rotation $\delta \alpha$ we get a compact product between inverse-variance filtered and Wiener-filtered Stokes parameters:
\begin{multline}
\widehat{\delta \alpha} \approx
   2  R^{-1} \begin{bmatrix} \bar{Q}  \\ \bar{U} \end{bmatrix} \begin{bmatrix}
        0 & -1 \\
        1 & 0 & 
    \end{bmatrix}  \begin{bmatrix} Q^ {\rm{WF}}  \\ U^ {\rm{WF}} \end{bmatrix} \propto \\ \propto -2(\bar{Q}U^ {\rm{WF}}-\bar{U}Q^ {\rm{WF}})\ .
\end{multline}

So far, we have yet to detect anisotropic birefringence, with the tightest constraints to date coming from ACT, SPT and BICEP 3-Keck data \cite{namikawaAtacamaCosmologyTelescope2020, bianchiniSearchingAnisotropicCosmic2020, collaborationBICEPKeckXVII2023}.

\subsubsection{Deflection from CMB gravitational lensing}

Large-scale structure (LSS) along the line of sight from the last scattering surface distorts CMB photons. Understanding this effect of CMB lensing allows us to constrain the amplitude of density fluctuations, neutrino masses, and dark energy (e.g. \cite{sherwinEvidenceDarkEnergy2011, Madhavacheril_2024}).

We can model CMB lensing through a remapping operation \cite{lewisWeakGravitationalLensing2006}

\begin{equation}
    X_{L}(\hat{n}) = O_{\mathrm{len}}(\vec{d})X \equiv  X(\hat{n}+\vec{d}(\hat{n}))\ ,
\end{equation}
where $\vec{d}$ is a deflection field. 

The derivation of the CMB lensing estimator have already been extensively presented in the literature, and we show only the final expression for the QE from CMB polarization only data \cite{huMassReconstructionCMB2002, Carron_2017, planckcollaborationPlanck2018Results2020}:


\begin{equation}
    \widehat{\vec{d}}^{\mathrm{QE}} =R^{-1}\bar{X} \cdot \vec{\nabla} X^{\rm{WF}}\ .
\end{equation}

The deflection field can be decomposed into gradient ($\widehat{\phi}$) and curl ($\widehat{\Omega}$) components \cite{Namikawa_2012}. The first, causes dilation and shear effects in the observed CMB, and is related to the linear order matter density fluctuations along the line of sight \cite{Schaan_2019}. So far, it has provided with an amazing opportunity to probe the late-time evolution of structure, thanks to high-significance detections in auto-spectrum \cite{planckcollaborationPlanck2018Results2020, Madhavacheril_2024, geCosmologyCMBLensing2024}, and in combination with large-scale structure (e.g. \cite{Farren_2024, qu2024atacamacosmologytelescopedr6}).

The latter is second order in density fluctuations, arising due to post-Born corrections \cite{Pratten_2016} and generating shear effects and tiny image-rotations of the CMB polarization \cite{carronDetectingRotationLensing2025c}. While it has yet to be detected, cross-correlations with large-structure look promising \cite{robertson2024detectlensingrotation}.\footnote{A curl component can also be generated by tensor perturbations \cite{coorayCosmicShearMicrowave2005, Namikawa_2012} or some systematic effect.}

\section{Reconstruction in the presence of multiple fields \label{sec:firstexperiments}}

\subsection{Experimental setup, simulations and codes}

We examine the performance of the QE standard estimator using full-sky simulations, avoiding mask-induced mean-field complications \cite{Benoit_L_vy_2013, namikawaBiashardenedCMBLensing2013} to focus on the fundamental benefits of likelihood methods.

Our simulations use a Planck 2013 cosmology with massless neutrinos \cite{carboneDEMNUniISWReesSciama2016a}. We model early-time cosmic birefringence through a Chern-Simons coupling between axion-like particles and photons, with a scale-invariant power spectrum $\frac{\left(L(L+1)\right)^{n_s}}{2\pi}C_L^{\alpha\alpha} = A_{CB}$ ($n_s=1$, $A_{CB} = 10^{-7}$, the expected $1\sigma$ limit for a CMB-S3-like survey \cite{pogosianFutureCMBConstraints2019}). For patchy reionization, we adopt the halo-based model from \cite{bianchiniInferenceGravitationalLensing2023}, assuming an extended reionization period ($\Delta z = 4$) with large bubbles. As our goal is a first validation of the pipeline, we assume zero cross-correlation between $\phi$ and $\tau$, and do not explore alternative reionization models.

We consider three experimental configurations: a CMB-S4-like survey ($1\ \mu \mathrm{K}\mathrm{-arcmin}$ noise, $1\ \mathrm{arcmin}$ beam) \cite{abazajianCMBS4ScienceBook2016a}, an SO-like survey ($6\ \mu \mathrm{K}\mathrm{-arcmin}$), and an SPT-3G-like survey ($1.6\ \mu \mathrm{K}\mathrm{-arcmin}$), both with $1.4\ \mathrm{arcmin}$ beams.\footnote{The SPT-3G-like case matches noise levels from SPT-3G Main data \cite{prabhu2024testingmathbflambdacdmcosmologicalmodel}.} For reconstruction, we use CMB $E$ and $B$ modes in the range $l \in [30,4000]$, targeting multipoles $L \in [2, 5000]$. We employ modified versions of \texttt{plancklens} \cite{planckcollaborationPlanck2018Results2020},\footnote{\url{https://github.com/carronj/plancklens/}} and \texttt{delensalot}, using gradient-lensed spectra in the QE response \cite{Fabbian_2019}.\footnote{Rotation and screening effects in filters are negligible compared to using lensed-only spectra \cite{namikawaTestingParityviolatingPhysics2019}} The code used to produce the results in this paper can be found in \texttt{jointmap}.\footnote{\url{https://github.com/Saladino93/jointmap}.}

\subsection{Reconstructing birefringence in the presence of CMB lensing \label{sec:cosmicrotlensing}}

As a starting point, we will focus on the joint presence of a primordial cosmic rotation, followed by gradient only CMB lensing. The observed CMB follows:

\begin{equation}
    X^{\mathrm{obs}} = \Beam\Big(e^{2i\alpha}X\Big)(\hat{n}+\vec{\nabla}\phi)+n \ , \label{eq:lensedalpha}
\end{equation}
where $n$ is some white noise and $\Beam$ is the beam.

\begin{figure}
    \centering
    \includegraphics[width=\linewidth]{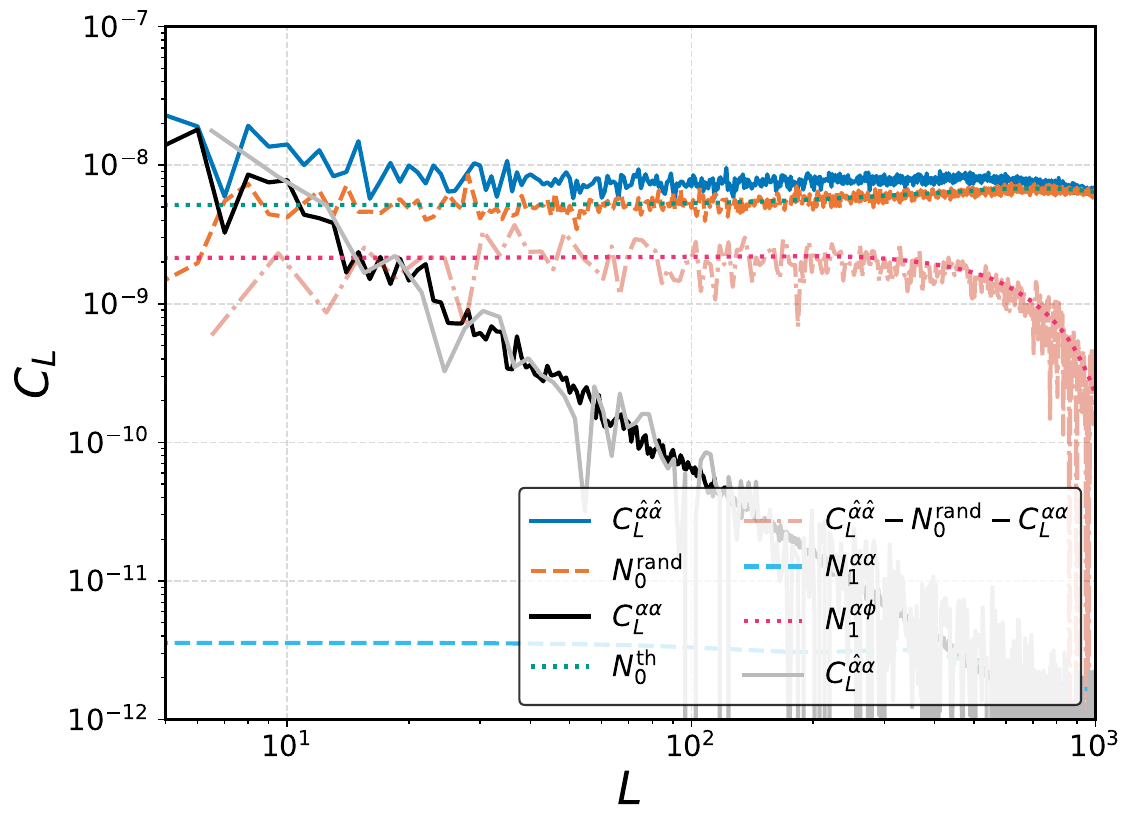}
    \caption{Angular power spectrum of the QE reconstructed cosmic rotation. In blue we show the total power spectrum \ref{eq:totalps}. In orange, the noise due to Gaussian contractions. In green, its theory prediction. In black, the input power spectrum, and in grey the cross-spectrum between the simulation and the input (binned spectrum). In light red and red we show the $N^{1,\alpha\phi}_L$ bias due to the presence of CMB lensing, from (binned spectrum) simulations and theory, respectively. Finally, in light blue, we show the theory $N_1^{\alpha\alpha}$ bias due to cosmic rotation power spectrum.}
    \label{fig:alphaqe}

    \footnotetext{The $N^{0}$ disconnected bias is computed by phase-randomizing each QE leg with $X_a^{j} = X_ae^{i\theta_j}$, $\theta_j\in[0,2\pi[$ 
    and calculating the QE power spectrum from a single simulation \cite{Das_2011}.}

\end{figure}

On the top panel of Figure \ref{fig:alphaqe} we show QE birefringence reconstruction results for one single simulation for the CMB-S4-like case. We show the dominant noise bias coming from Gaussian contractions (orange), and limited by the lensed CMB.  The $N_1^{\hat{\alpha}\phi}$ bias (red) induced by the CMB lensing potential $\phi$, that needs to be subtracted to give accurate results \cite{namikawaTestingParityviolatingPhysics2019}. We can clearly see that this overwhelms the standard $N_1^{\alpha\alpha}$ bias (light blue), as highlighted in the bottom panel of Figure \ref{fig:alphaqe}.

While bias-hardening typically addresses contamination, as explained it proves insufficient for the $N^{\hat{\alpha},\phi}_1$ bias.\footnote{Even if bias-hardening would have worked for mitigating $N_1$ biases, at linear order, and for small sources of anisotropy, the gradient mode of lensing and cosmic rotation are orthogonal \cite{kamionkowskiHowDeRotateCosmic2009, gluscevicDeRotationCosmicMicrowave2009,namikawaTestingParityviolatingPhysics2019, caiImpactAnisotropicBirefringence2023b}, precluding standard bias-hardening.} However, delensing offers a solution: removing lensing effects before estimating cosmic rotation can mitigate both the $N^{\hat{\alpha},\phi}_1$ bias and reduce estimator noise \cite{carronDetectingRotationLensing2025c}. This is easily understood from our Equation \ref{eq:lensedalpha}. With perfect delensing using an estimated potential $\widehat{\phi}$, the lensed and rotated signal $S(\hat{n}) = \Big(e^{2i\alpha}X\Big)(\hat{n}+\vec{\nabla}\phi)$ becomes $S^{\mathrm{del}} \equiv \Big(e^{2i\alpha}X\Big)(\hat{n}+\vec{\nabla}\phi-\vec{\nabla}\widehat{\phi}) \approx \Big(e^{2i\alpha}X\Big)(\hat{n})$, directly revealing the rotated primordial CMB. Maximum a posteriori methods \cite{hirataAnalyzingWeakLensing2003, Carron_2017} provide a systematic approach to this delensing strategy, which we develop in the next section.

\section{Joint MAP reconstruction \label{sec:jointmaprec}}

While the quadratic estimator in Section \ref{sec:distortingphysics} emerges from a single Newton-Raphson iteration of the likelihood, we can improve reconstruction by performing multiple iterations (defining the QE as $\mathrm{itr}=0$)
\begin{equation}
    \widehat{\xi}^{\mathrm{itr+1}} = \widehat{\xi}^{\mathrm{itr}}-H_{\mathrm{itr+1}} \frac{\delta L}{\delta \xi}\Big\vert_{\xi=\widehat{\xi}^{\mathrm{itr}}}  \ ,
\end{equation}
and incorporating prior information. This forms the basis of maximum a posteriori (MAP) estimation \cite{hirataAnalyzingWeakLensing2003, hirataReconstructionLensingCosmic2003, Carron_2017, milleaBayesianDelensingDelight2020}. Building on techniques developed for CMB lensing marginal MAP  \cite{Carron_2017}, we extend this framework to jointly estimate multiple distortion fields.\footnote{The CMB lensing marginal MAP integrates out the primordial CMB to give a probability density function of CMB lensing given the data to look for the most probable CMB lensing field \cite{carron2025likelihoodbasedanalysisgravitationallydelensed}.}

Consider $N$ CMB distorting fields $\xi \equiv (\xi_1, ..., \xi_N)$ acting sequentially through operators $O_i$:
\begin{equation}
    X^{\rm dat} = \Beam O_1 O_2 ... O_N Y X + f + n,
\end{equation}
with $X$ the primordial CMB, $Y$ the spherical harmonic synthesis, $\Beam$ a linear response matrix that includes the beam, $n$ the instrumental noise, uncorrelated with any astrophysical component, and $f$ any other source of CMB statistical anisotropy, such as foregrounds, that from now on we will set this to 0.\footnote{Though similar techniques can be used for these other sources.}

The MAP estimator maximizes the posterior:
\begin{multline}
    \ln p(\xi_1, \xi_2, \xi_3, ...|X^{\rm{dat}}) =  \ln p(X^{\rm{dat}}|\xi_1, \xi_2, \xi_3, ...) + \\  \ln p_{\xi_1, \xi_2, \xi_3, ...}(\xi_1, \xi_2, \xi_3, ...)\ + \mathrm{const} , \label{eq:lnposterior}
\end{multline}
given a log-likelihood $\ln p(X^{\rm{dat}}|\xi_1, \xi_2, \xi_3, ...)$ and a log-prior $\ln p_{\xi_1, \xi_2, \xi_3, ...}(\xi_1, \xi_2, \xi_3, ...)$. To maximize this we will need to get its total gradient, that can be broken into three parts\cite{Carron_2017}:

\begin{equation}
g_{\mathrm{tot}}=g_{\mathrm{QD}}^{\xi}-g_{\mathrm{MF}}^{\xi}+g_{\mathrm{PR}}^{\xi}=0\ ,\label{eq:totlngradient}
\end{equation}
a quadratic part, a mean-field term, and a prior term. Let's start with the quadratic part.

\subsection{Quadratic part of the likelihood}

We consider a Gaussian log-likelihood for the data \cite{Carron_2017}

\begin{multline}
    \ln p(X^{\mathrm{dat}}|\xi_1, \xi_2, \xi_3, ...) = -\frac{1}{2}X^{\rm{dat}} \cdot \mathrm{Cov}_{\xi_1, \xi_2, \xi_3, ...}^{-1} X^{\rm{dat}} + \\ - \frac{1}{2}\det \mathrm{Cov}_{\xi_1, \xi_2, \xi_3, ...}+ \rm const\ \ ,
\end{multline}
where the first addend on the right hand side of the equation is the quadratic part, and the second addend is the determinant of the covariance matrix:

\begin{align}
    \mathrm{Cov}_{\xi_1, \xi_2, \xi_3, ...} = \langle X^{\mathrm{dat}}X^{\mathrm{dat},\dagger}\rangle &= \nonumber \\
    &= \Beam M Y C^{\mathrm{unl}}Y^{\dagger}M^{\dagger}\Beam^{\dagger} + N \label{eq:covariance}
\end{align}
where $C^{\rm unl}$, $N$ are the covariance for the unlensed CMB and the noise, respectively, and 
$M = \prod_{i=1}^{N} O_i$ is the full forward operator, while $M^{\dagger} = \prod_{i=N}^{1} O_i^{\dagger}$ the full adjoint operator. In particular, we explicitly specify that the covariance has a structure that depends on the realization of the different components $\xi=(\xi_1, \xi_2, \xi_3, ...)$ . 

To maximize the posterior \ref{eq:lnposterior}, we need to take the functional derivatives with respect to each $\xi_{j}$. This involves calculating:

\begin{multline}
\frac{\delta \mathrm{Cov}^{-1}_{\xi_1, \xi_2, \xi_3, ...}}{\delta \xi_{j}} = \\ - \mathrm{Cov}^{-1}_{\xi_1, \xi_2, \xi_3, ...} \frac{\delta \mathrm{Cov}_{\xi_1, \xi_2, \xi_3, ...}}{\delta \xi_{j}}  \mathrm{Cov}^{-1}_{\xi_1, \xi_2, \xi_3, ...}
\end{multline}
The quadratic part of the likelihood gradient is then:

\begin{equation}
     \frac{1}{2}\overline{X}^d \frac{\delta \mathrm{Cov}_{\xi_1, \xi_2, \xi_3, ...}}{\delta \xi_{j}}  \overline{X}^d\ ,
\end{equation}
with an inverse variance filtered CMB map $\bar{X}^d \equiv  \mathrm{Cov}^{-1}_{\xi_1, \xi_2, \xi_3, ...} X^d$.
From Equation \ref{eq:covariance}, the derivative of the covariance with respect to the field $\xi_{j}$ involves

\begin{multline}
    \frac{\delta \mathrm{Cov}_{\xi_1, \xi_2, \xi_3, ...}}{\delta \xi_{j}} =  \Beam \frac{\delta M}{\delta \xi_j} C^{\mathrm{unl}}M^{\dagger}\Beam^{\dagger} + \mathrm{c.c.}
\end{multline}
But, assuming operator $O_j$ depends only on the corresponding $\xi_j$,\footnote{In principle, we should consider how a field, e.g. CMB lensing, affects the action of other operators, e.g. patchy tau. We ignore this for now, and we defer to the future checking the impact of our choice here.}

\begin{equation}
    \frac{\delta M}{\delta \xi_j}  = ... O_{j-1} \frac{\delta O_j} {\delta \xi_j} O_{j+1} ...
\end{equation}
Therefore, the quadratic part (QD) is:

\begin{equation}
    g_{\mathrm{QD}}^{\xi_j}(\hat{n}) = \\  \Big(\Beam\overline{X}^d \Big) \cdot \Big(... O_{j-1} \frac{\delta O_j} {\delta \xi_j} O_{j+1} ... X^{WF}\Big)(\hat{n})\ . \label{eq:qdpart_explicit}
\end{equation}
In this expression, the first factor between parentheses in the product is an inverse variance map, and the second factor
\begin{equation}
X^{WF}=C^{\mathrm{unl}}M^{\dagger}\Beam^{\dagger} \overline{X}^d,
\end{equation}
a Wiener filtered CMB map looking for the primordial CMB to which we apply any physics acting on it before and after the action of $\xi_j$. The quadratic part looks only at the variation in the covariance introduced by the relevant distortion. This expression is basically the QE of Equation \ref{eq:likeqe}, but evaluated based on our knowledge of the fields.


For our numerical implementation, following \cite{Carron_2017} we will get the inverse variance filtered observed CMB from
\begin{equation}
   \bar{X} =  \Beam^{\dagger} \overline{X}^d = \Beam^{\dagger} N^{-1} \Big[X^d-\Beam M Y X^{\rm{WF}}\Big]\ ,
\end{equation}
and the Wiener filtered CMB from

\begin{align}
X^{\rm{WF}}= \Big[ (C^{\mathrm{unl}})^{-1} + Y^{\dagger}M^{\dagger} \Beam^{\dagger} N^{-1} \Beam M Y \Big]^{-1} \times \nonumber \\
  \times Y^{\dagger} M^{\dagger} \Beam^{\dagger} N^{-1} X^{d}\ .
\end{align}
Once we get the gradient part $g_{\mathrm{QD}}^{\xi_j}$ for each source of CMB anisotropy $\xi_j$.

\subsection{The mean-field and the prior}

To complete our gradient calculation, we need the other two components: the mean-field and the prior, both ignored for the QE.

The mean-field term can be crucial for MAP reconstruction. It captures anisotropy sources beyond our targeted distortions, including effects from survey masks, and new anisotropies induced by “de-operating” the noise \cite{carron2025likelihoodbasedanalysisgravitationallydelensed}. 

We estimate a MAP mean-field as \cite{hansonWeakLensingCMB2010, Carron_2017}: 
\begin{equation}
    g_{\mathrm{MF}}^{\xi_j} = \langle g_{\mathrm{QD}}^{\xi} \rangle
\end{equation}
where we average over the data realizations, while keeping the distortion fields $\xi$ at their MAP solution $\xi_{\mathrm{MAP}}$. In Appendix \ref{sec:appendixmf} we show its relative magnitude for the fields considered here.

Finally, we include prior information about the distortion fields. For simplicity, we assume a Gaussian log-prior on all of our components, taking into account possible correlations among fields through a covariance $C_{\mathrm{PR}}$ (ignoring the covariance determinant as we care about gradients):

\begin{equation}
    \ln p_{\xi_1,\xi_2,\xi_3,...} = -\frac{1}{2}\xi^{\dagger}C_{\mathrm{PR}}^{-1}\xi\ ,
\end{equation}
with a gradient for the field $\xi_j$

\begin{equation}
    g_{\mathrm{PR}}^{\xi_j} = -C_{\mathrm{PR}}^{-1}\xi=-\sum_i(C_{\mathrm{PR}}^{-1})_{ij}\xi_{j}\ ,
\end{equation}
though in principle we can account for non-Gaussian priors \cite{darwish2024nongaussiandeflectionsiterativeoptimal}. This prior will filter out noisy modes from our reconstruction, effectively leading to a Wiener filter \cite{legrandLensingPowerSpectrum2022}. If the mean-field can be ignored, then the prior becomes also a diagnostic tool, as at the MAP point $g_{\mathrm{QD}}+g_{\mathrm{PR}}=0$.

\begin{figure*}[!ht]
    \centering  \includegraphics[width=\linewidth]{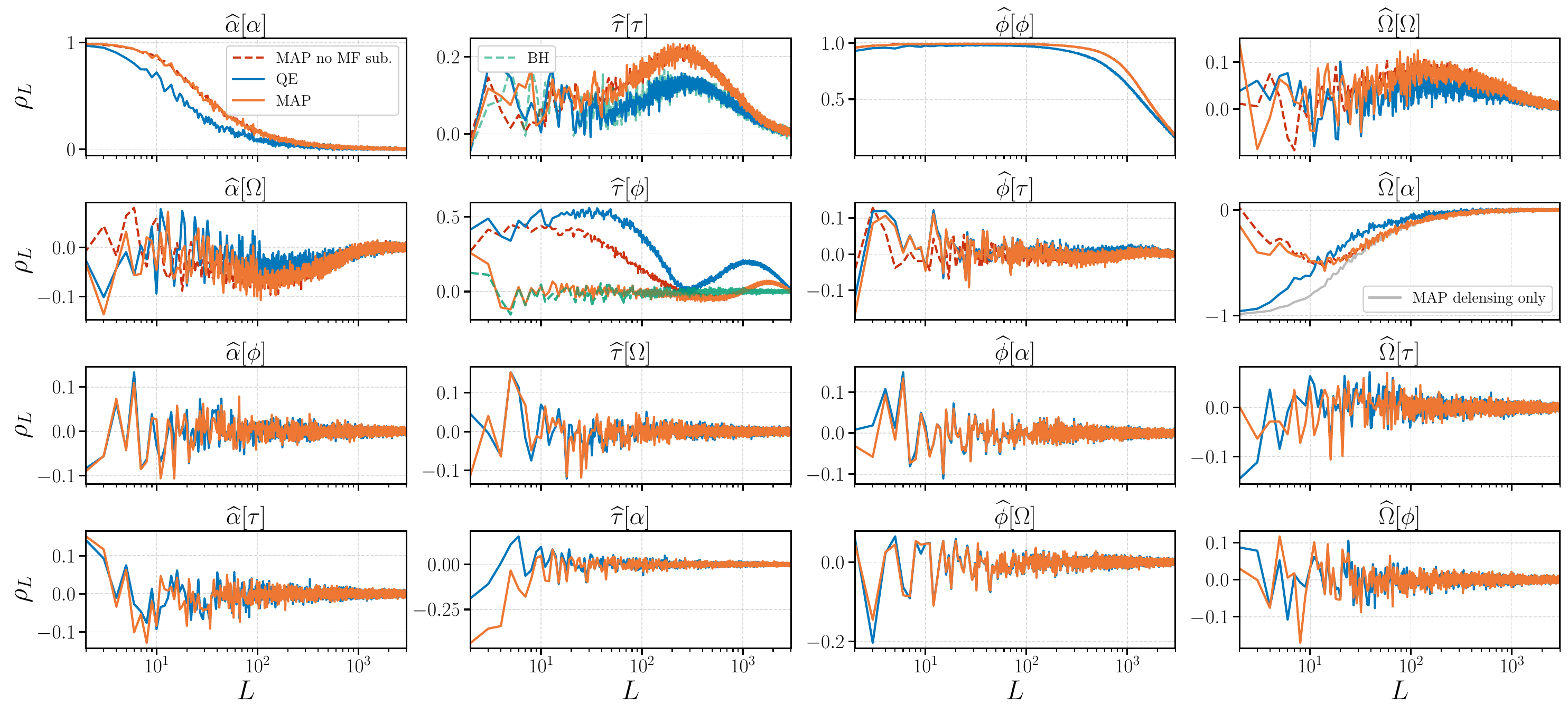}
    \caption{Cross-correlation coefficients between reconstructed estimators and input fields for a CMB-S4-like experiment. \textbf{Organization}: Each column shows results for a different estimator $\widehat{x} \in \{\widehat{\alpha},\widehat{\tau},\widehat{\phi},\widehat{\Omega}\}$ (cosmic rotation, patchy reionization, lensing gradient, and lensing curl, respectively). Each row shows how a given estimator correlates with different input fields $x \in \{\alpha,\tau,\phi,\Omega \}$. We remind that we make the incorrect but useful assumption that $\tau$ and $\phi$ are uncorrelated, for testing purposes.  \textbf{Main results}: Blue curves show standard quadratic estimator (QE) performance, while orange curves show our MAP approach. Results are averaged over 16 simulations. (1) The top row demonstrates that MAP improves correlation with the corresponding input field across all estimators.  (2) The second row reveals that MAP reduces contamination from other fields, particularly for patchy tau (second panel, less contaminated by lensing) and curl mode (fourth panel, less contaminated by cosmic birefringence). (3) Dashed-red lines show MAP without mean-field correction, highlighting its importance for unbiased tau reconstruction. Green lines show standard bias-hardening (BH) for comparison. Grey lines in the curl estimator panel show results when only delensing (without de-rotation) is applied. The two bottom rows show combinations of different estimators.} 
    \label{fig:xcorrcoeffs}
\end{figure*}

\section{Results \label{sec:results}}

Having established the theoretical framework, we now apply our code to demonstrate its practical effectiveness. We find the maximum a posteriori solution of Equation \ref{eq:totlngradient} through an iterative approach. We initialize our iterations with a Wiener-filtered quadratic estimator (QE) estimate that suppresses noisy modes\cite{Carron_2017}:
\begin{equation}
    \mathbf{W} = \Big[\mathbf{C}^{-1}+\mathbf{R}\Big]^{-1}\mathbf{R},
\end{equation}
where $\mathbf{C},\mathbf{R}$ are the signal and response matrices, respectively, accounting for all possible cross-correlations between fields. The iterative solution employs 15 Newton-Raphson-like steps,\footnote{In particular, we use the Broyden-Fletcher-Goldfarb-Shanno (BFGS) algorithm.} as additional iterations yield minimal improvements for our goals described here. For CMB Wiener filtering we use a conjugate gradient descent with $10^{-6}$ convergence tolerance. We use a uniform step size $\lambda = 0.5$ between iterations.\footnote{Though potentially better behaved steppers could be used, such as one that separates large and small scales, to achieve improved convergence properties, we chose this simpler approach. We could also have run for more iterations (30), as this ensures we can remove residual gradients. This would introduce more computational power, unneeded for our presentation of the code.}

\subsection{Joint reconstruction of cosmic birefringence, CMB lensing, and patchy tau}

For our initial demonstration, we examine a scenario where the CMB undergoes sequential distortions: early-time polarization rotation $O_{\mathrm{rot}}(\alpha)$, amplitude modulation $O_{\mathrm{patchy}}(\tau)$, and deflection $O_{\mathrm{len}}(\vec{d})$, resulting in a forward operator $M = O_{\mathrm{len}}O_{\mathrm{patchy}}O_{\mathrm{rot}}$. For testing purposes, we assume uncorrelated $\tau$ and $\phi$, even though they are physically related as they originate from the same density field. We obtain the mean-field gradient with a quick approximation presented in an Appendix of \cite{Carron_2017}, using a limited set of five simulations. While in practice better approximations for the mean-field can be used, we do this for computational speed.

Figure \ref{fig:xcorrcoeffs} shows cross-correlation coefficients of the reconstructed fields for a CMB-S4-like configuration, calculated as
\begin{equation}
    \rho_L = \frac{\langle \widehat{x} y \rangle}{\sqrt{\langle \widehat{x} \widehat{x} \rangle\langle yy \rangle}}
\end{equation}
where $\widehat{x} \in \{\widehat{\alpha},\widehat{\tau},\widehat{\phi},\widehat{\Omega}\}$ is a reconstructed field, and $y \in \{\alpha,\tau,\phi,\Omega \}$ an input field used to distort the primordial CMB simulation.

First, examining the top row (where $y=x$), we see that the MAP estimator (orange) consistently improves reconstruction fidelity across all fields compared to the standard QE (blue). This improvement stems primarily from reductions in both the $N_0$ and $N_1$ biases.

\begin{figure*}[ht!]
    \centering
    \includegraphics[width=\linewidth]{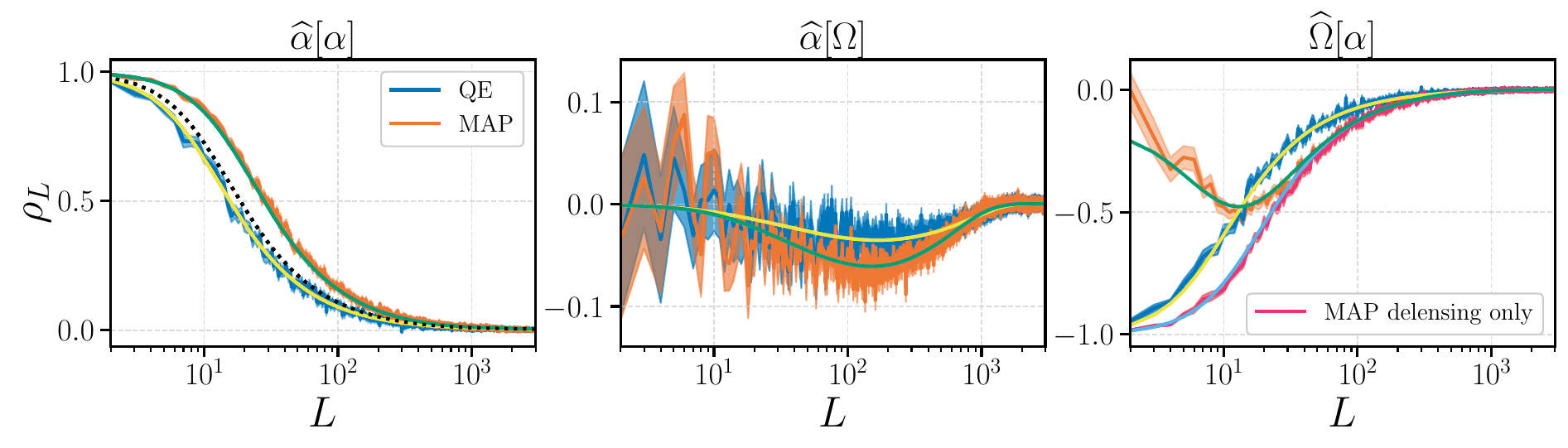}
    \caption{Predicted cross-correlation coefficients for the cosmic birefringence and curl estimators $\widehat{\alpha}, \widehat{\Omega}$ with the inputs $\alpha$, $\Omega$ for our CMB-S4-like setup. In yellow QE predictions, in green MAP, light blue MAP with no derotation. On the left, for the QE we show in the dotted black line a small mismatch with simulations when excluding the $N^{\widehat{\alpha}\phi}_1$ term. The shaded areas are scatter from sims.}
    \label{fig:theoryxcorr}
\end{figure*}

We test this for the more relevant case here of cosmic birefringence. The predicted cross-correlation coefficient between $\widehat{\alpha}$ and the input $\alpha$ is:
\begin{equation} \rho^{\mathrm{th},\widehat{\alpha}}[\alpha]=\sqrt{\frac{C_L^{\alpha\alpha}}{C_L^{\alpha\alpha}+N_0+N_1^{\widehat{\alpha}\alpha}+N_1^{\widehat{\alpha}\phi}}}\ ,
\end{equation}
where we follow the recipe described in Appendix \ref{sec:appendixnoise} to get the noise curves, and ignore $N_1^{\widehat{\alpha}\Omega}, N_1^{\widehat{\alpha}\tau}$. We show predictions in Figure \ref{fig:theoryxcorr}, left panel. In particular, the omission of $N_1^{\widehat{\alpha}\phi}$ for the QE reduces the accuracy of our prediction, while the MAP approach is more robust to small mis-modellings.

Second, the middle rows reveal how each estimator cross-correlates with other input fields, highlighting contamination effects. The patchy tau MAP estimator (orange line, second column) shows robustness against CMB lensing potential contamination, performing similarly to the bias-hardened (BH) estimator (green) but with the added advantage of stronger correlation with the true $\tau$ signal (first row, second panel from the left). Importantly, our tests demonstrate the critical role of the mean-field correction. Without it (dashed-red lines), we still observe increased cross-correlation with the input (first row, second panel) but limited improvement in reducing contamination (second row, second panel from the left). This confirms that proper bias-hardening requires accounting for the full likelihood gradient, including both quadratic and mean-field terms. Finally, the CMB lensing curl estimator presents a special challenge due to its high sensitivity to cosmic birefringence—both effects create rotation-like patterns in the CMB. Delensing alone (without de-rotation) increases contamination from cosmic birefringence by better revealing the rotated CMB signal. By jointly de-rotating with our estimated $\hat{\alpha}$, we partially mitigate this effect, particularly at large scales where the signal-to-noise is highest.

We can explain simulation results by looking at theory cross-correlation coefficients for this case, calculated as
\begin{widetext}
\begin{equation}
\rho^{\mathrm{th},\widehat{\Omega}}[\alpha]= \frac{\frac{R^{\widehat{\Omega}\alpha}}{R^{\widehat{\Omega}\Omega}}C_L^{\alpha_{\mathrm{res}}\alpha}}{\sqrt{\Big[C_L^{\Omega\Omega}+   \Big(\frac{R^{\widehat{\Omega}\alpha}}{R^{\widehat{\Omega}\Omega}}\Big)^2C_L^{\alpha_{\mathrm{res}}\alpha_{\mathrm{res}}}+N_0+N_1^{\widehat{\Omega}\alpha}+N_1^{\widehat{\Omega}\phi}+N_1^{\widehat{\Omega}\Omega}\Big]C_L^{\alpha\alpha}}}\ ,
\label{eq:crosscorrcontaminant}
\end{equation}
\end{widetext}
where the responses $R$ and $N_1$ biases are calculated using partially delensed spectra, and $C_L^{\alpha_{\mathrm{res}}\alpha_{\mathrm{res}}}$ is the residual signal power spectrum from cosmic birefringence, as calculated in Appendix \ref{sec:appendixnoise}. This shows that without derotating the CMB spectra we are able to get good matching with simulations.

Finally, in principle we could QE bias-harden the curl estimator against the rotation estimator. But it proves not convenient due to the high correlation between $\widehat{\alpha}$ and $\widehat{\Omega}$ estimators \cite{carronDetectingRotationLensing2025c}; the resulting noise penalty from bias-hardening scales as $\frac{1}{1-r^2}$, where $r$ is the cross-correlation coefficient between the two $\widehat{\alpha}$ and $\widehat{\Omega}$ estimators \cite{Darwish_2021}.\footnote{Similarly, also the $\hat{\alpha}$ estimator is affected by $\Omega$, so in this case the impact is smaller. We see some small cross-correlation coefficients in Figure \ref{fig:xcorrcoeffs} increase in the $\widehat{\alpha}$ estimator with $\Omega$, as well as $\widehat{\tau}$ estimator with $\alpha$. As these biases are not on relevant scales of analysis, where SNR of estimators is low, we decided to investigate this in the future, though with better mean-field subtraction and stepping criteria we can improve our results.} 

\begin{figure}
    \centering
    \includegraphics[width=\linewidth]{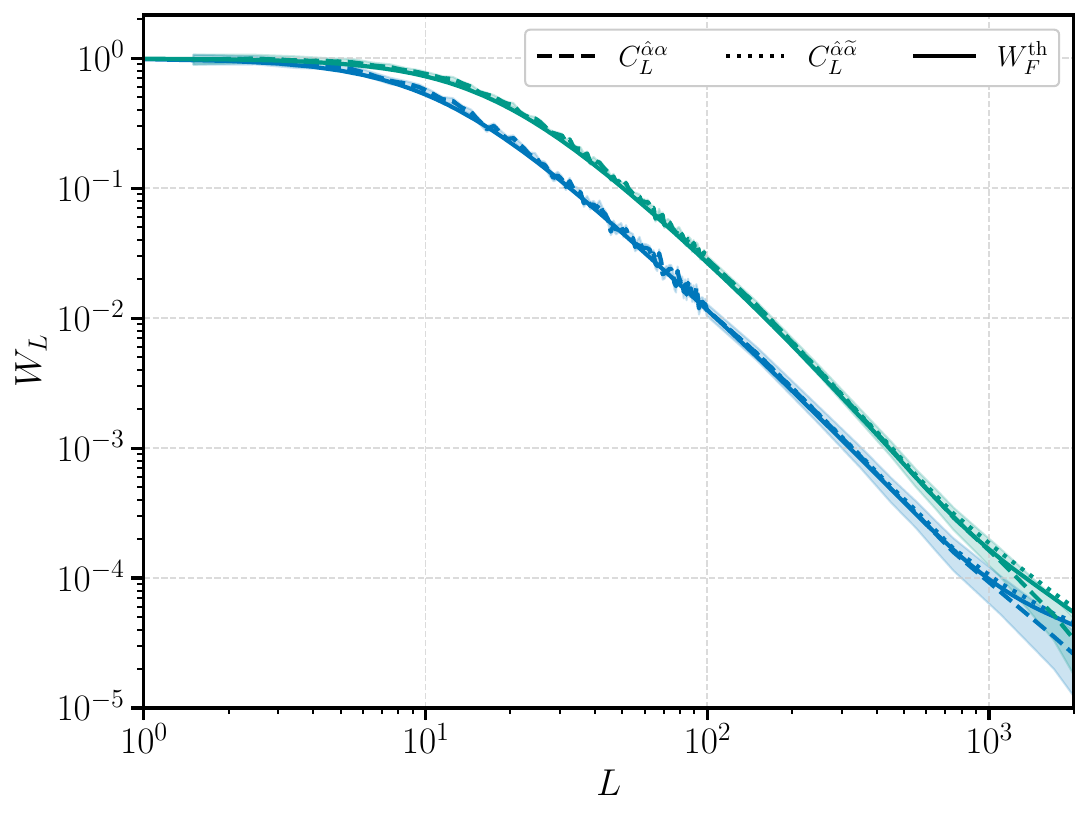}
    \caption{Wiener filter across iterations. In blue we show the QE, and in green the MAP. Solid lines are theory calculations. Dashed lines are the reconstructed fields cross the input. And finally, dotted lines are reconstructed fields cross lensed-input. }
    \label{fig:wienerfilter}
\end{figure}

\begin{figure*}[!htp]
\centering
\begin{minipage}[t]{0.49\textwidth}
  \centering
  \includegraphics[width=\linewidth]{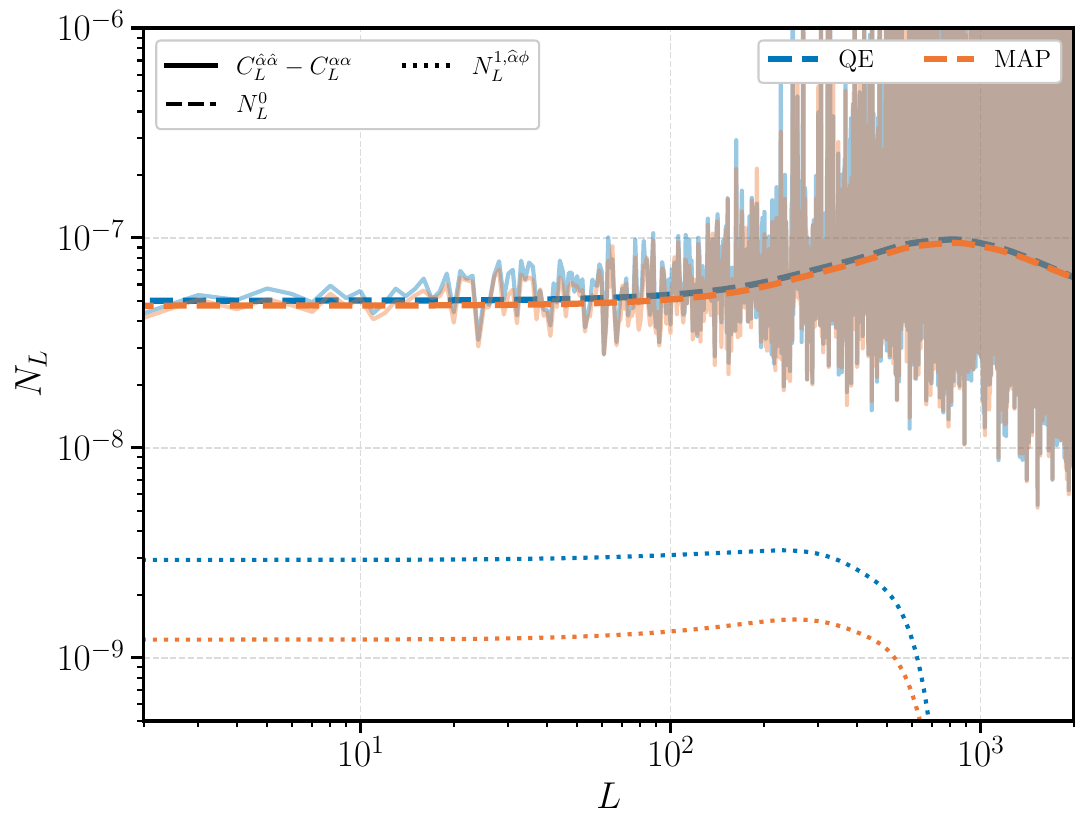}
  \vspace{-0.5cm}  
  \textbf{(a)} SO-like
\end{minipage}%
\begin{minipage}[t]{0.49\textwidth}
  \centering
  \includegraphics[width=\linewidth]{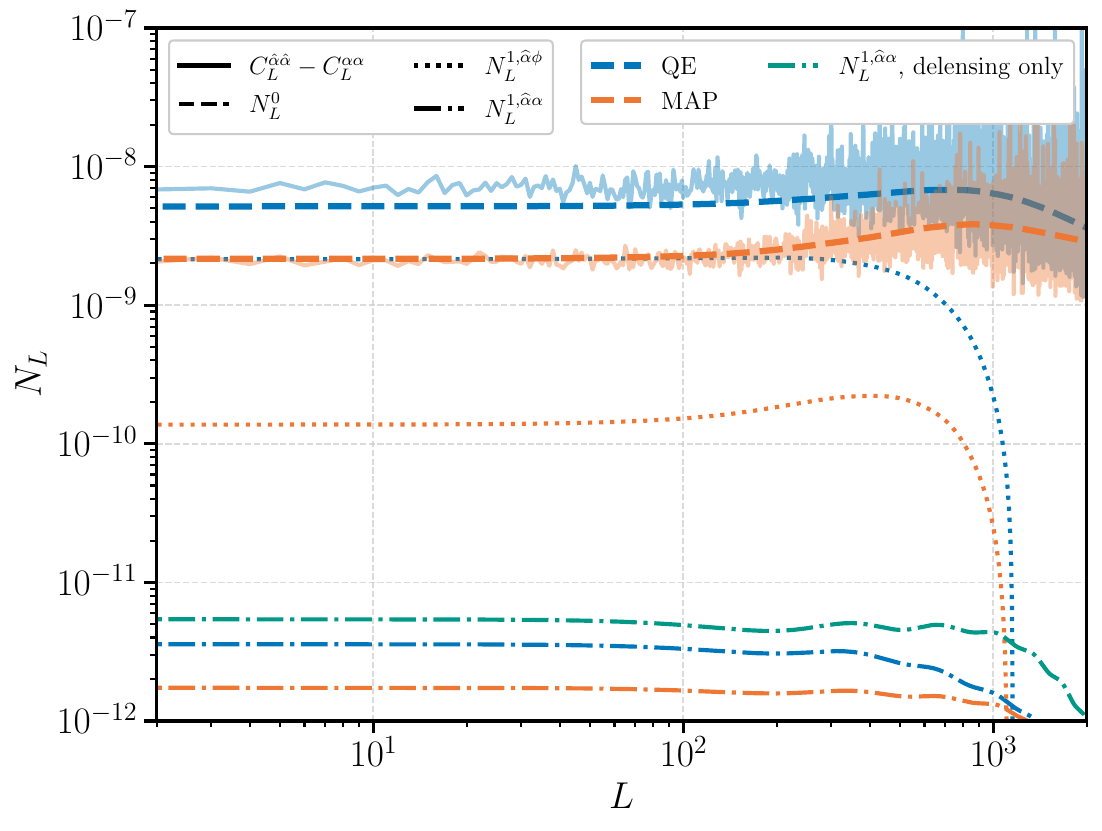}
  \vspace{-0.5cm}  
  \textbf{(b)} CMB-S4-like
\end{minipage}

\vspace{0.5cm}  
\begin{minipage}[t]{0.49\textwidth}
  \centering
  \includegraphics[width=\linewidth]{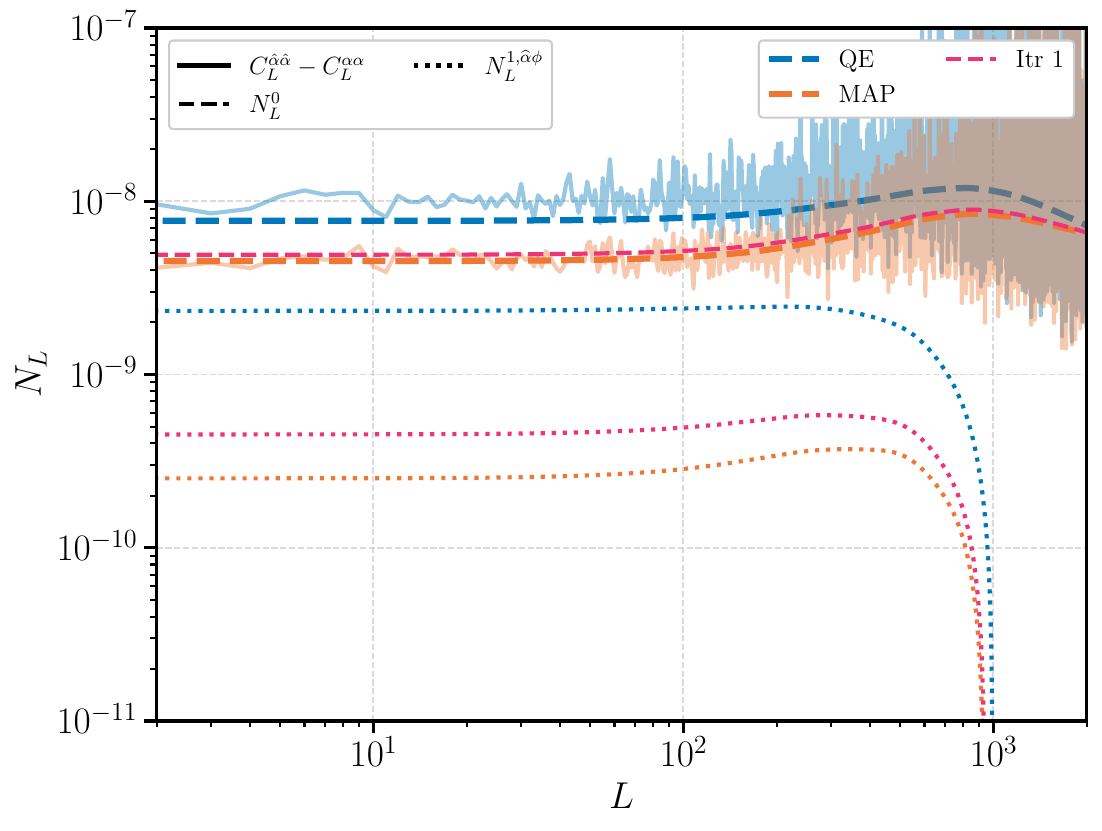}
  \vspace{-0.5cm}  
  \textbf{(c)} SPT-3G-like
\end{minipage}
\vspace{0.5cm}
\caption{Raw noise $N_0$ spectra with and without delensing and derotating. In solid, we show simulations results, using an empirical Wiener filter. In dashed, analytical $N_0$ based on the simple prescription described in the text. The dotted, dotted-dashed lines are analytical $N_1^{\hat{\alpha},\phi}$, $N_1^{\hat{\alpha},\alpha}$ calculations, respectively. On the top-right, we have results for a CMB-S4-like survey. We can see that $N_0$ decreases by a factor of around 2.5 at large multipoles ($L<100$), and by a larger factor is the decrease in $N_1^{\hat{\alpha},\phi}$ (blue vs orange curves). Delensing only leads to similar noise improvements, and it increases the $N_1^{\hat{\alpha},\alpha}$ bias (green curve). On the top-left, we show results for an SO-like survey. We can clearly see while in this case we do not gain much in $N_0$, we reduce by a factor of $\sim 2$ the $N_1^{\hat{\alpha},\phi}$ bias, showing that MAP methods will be useful even when there is not a big reduction in noise. Finally, on the bottom we show results for an SPT-3G-like survey, where we see again a large reduction in $N_1^{\hat{\alpha},\phi}$. We also show for a comparison the $N_1^{\hat{\alpha},\phi}$ after one single iteration, making standard delensing very useful.}
\label{fig:iterativerotation}
\end{figure*}

\subsection{Reconstructing birefringence in the presence of CMB lensing: the MAP version}

We now run the example of Section \ref{sec:cosmicrotlensing} using the MAP framework, where the forward operation $M = O_{\mathrm{len}}O_{\mathrm{rot}}$ is given by a rotation followed by a CMB lensing deflection. At each iteration, we jointly estimate both the lensing potential $\phi$ and rotation angle $\alpha$.

We normalize our results with the inverse of an empirical Wiener filter calculated as \cite{legrandLensingPowerSpectrum2022}

\begin{equation}
    W_L^{\mathrm{emp}} = \Big\langle \frac{C_L^{\widehat{\alpha}\alpha}}{C_L^{\alpha\alpha}}\Big\rangle_{\mathrm{sims}} \ ,
\end{equation}
to ensure that we can recover the input. Figure \ref{fig:wienerfilter} compares this empirical filter to a theoretical expectation $W_L^{\mathrm{th}}=C_L^{\alpha\alpha,\mathrm{th}}/(C_L^{\alpha\alpha,\mathrm{th}}+N_0)$. The agreement at large scales breaks down at small scales due to a subtle effect: our estimator reconstructs the lensed rotation field $\beta(\hat{n})=\alpha(\hat{n}+\vec{\nabla}\phi)$ rather than $\alpha$ itself. Cross-correlation with this lensed field restores agreement with theory.\footnote{The scale-invariant spectrum of our input $\alpha$ ensures large scales remain unaffected by lensing at lowest order \cite{lewisWeakGravitationalLensing2006}. Previous studies' practice of rotating simulations after lensing while studying primordial birefringence, while formally incorrect, is justified by the large-scale dominance of the signal-to-noise.}

From the total power spectrum, we estimate a residual power as:
\begin{equation}
    \Delta = C_L^{\hat{\alpha}\hat{\alpha}}-C_L^{\alpha\alpha}\ .
\end{equation}
We compare this quantity to raw noise curves, and to theory $N_1^{\widehat{\alpha}\phi}$ biases induced by CMB lensing. 
Figure \ref{fig:iterativerotation} presents our results. For the CMB-S4 and SPT-3G like cases, we show a substantial reduction in the noise $N_0$ and, indirectly, the $N_1^{\widehat{\alpha}\phi}$ bias after delensing and derotating. This can be seen by matching of the residual power with the theory noise. The noise reduction factors of 2.5 and 1.6 for CMB-S4 and SPT-3G respectively translate into improvements in the detection of the birefringence amplitude $A_{CB}$:\footnote{Assuming a Gaussian variance on the spectra. This is not entirely correct on large-scales where we have few modes. In principle we could use simulations, but for our exercise here of comparing relative improvements this should be fine.}
\begin{multline}
    \mathrm{SNR}^2(L_{\mathrm{max}}) \sim \\ \sim \sum_{L=1}^{L_{\mathrm{max}}}  (2L+1)f_{\rm{sky}}\frac{(C_L^{\alpha\alpha})^2}{[C_L^{\alpha\alpha}+N_0+N_1^{\hat{\alpha}\phi}]^2}\ ,
\end{multline}
where $f_{\rm{sky}}$ is the sky fraction of the survey, and $L_{\mathrm{max}}$ is the maximum multipole of the analysis. We find that going from QE to MAP, we find improvements in the SNR of $70\%$ and $40\%$, for CMB-S4/SPT-3G respectively. While for SO we get a mere $5\%$.\footnote{We ignore $N_1^{\alpha\alpha}$ in calculating these numbers. This can be a source of cosmological information, and so can be used as a signal \cite{namikawaTestingParityviolatingPhysics2019}. De-rotation will reduce it. But for the sake of this exercise, we are more interested into gauging the benefits of reducing the two largest biases, $N_0$ and $N_1^{\widehat{\alpha}\phi}$.}

For the reduction in the noise we also run the MAP framework using $\phi_{\rm{MAP}}$ only in the filtering, corresponding to delensing only.\footnote{In this case, even if there are $E$ and $B$ modes before lensing, we assume there are only $E$ modes, and therefore find a Wiener filtered estimate of these.} We do not see big differences in $N_0$, but Figure \ref{fig:iterativerotation}, panel $\boldsymbol{(b)}$ shows that in this case $N_1^{\alpha\alpha}$ increases. If treated as a bias this can be problematic. Derotation restores the proper MAP solution, and we see reductions in the $N_1^{\widehat{\alpha}\alpha}$ bias in this case (orange vs green curves in the Figure).

Notably, even a high-noise SO-like case that shows minimal $N_0$ improvements gains a factor of two reduction in lensing induced biases. The reason is that the $N_L^{\hat{\alpha}\phi}$ bias depends strongly on the residual lensing power $C_L^{\phi\phi}(1-\rho_L^2)$, where $\rho$ is the cross-correlation coefficient of the reconstruction with the true lensing field \cite{Carron_2017_internal}. This demonstrates the broad utility of our approach across different experimental regimes. Moreover, Figure \ref{fig:iterativerotation}, panel $\boldsymbol{(c)}$, shows that even a single delensing iteration substantially reduces the $N_1^{\hat{\alpha}\phi}$ for the SPT-3G-like survey. These observations suggest that current surveys can already benefit in reducing biases through simple delensing techniques (by combining CMB lensing internal reconstructions with external tracers, e.g. \cite{Sherwin_2015}), even if they do not benefit in terms of noise.

\section{Conclusions \label{sec:conclusions}}

The precise measurement of CMB distortion effects stands as a key objective for ground-based high-resolution surveys. While these distortions are typically analyzed independently, we demonstrate that joint analysis through likelihood-based methods can improve their reconstruction while naturally accounting for their mutual contamination; a key message of our work, is that current surveys can already benefit from a MAP approach by analyzing distortion fields with the delensed CMB. This has the consequence of more robust analyses.
A key point from our analysis was the crucial role of the mean-field correction in likelihood-based reconstruction, to preserve bias-hardening properties for the MAP solution. 
Going beyond, as for SPT-3G 2026 or CMB-S4, Our extension of the \texttt{delensalot} code enables simultaneous reconstruction of multiple CMB distorting fields, including lensing gradient and curl modes, cosmic birefringence, and patchy reionization. For these deep CMB surveys, the MAP approach delivers dual benefits: improved signal-to-noise and reduced biases. This proves particularly valuable for cosmic rotation reconstruction, where standard bias-hardening techniques struggle to address lensing induced contamination. 

We showed for the first time that even higher-noise surveys like SO can benefit from the MAP approach/simple delensing through reduced lensing bias, despite minimal improvements in reconstruction noise.

One of the key issues with our first application of our code, is the fact that the curl estimator can still be biased by the presence of (a hypothesized) early-time cosmic birefringence. The curl estimator is expected to give some detection in combination with large-scale structure in the next years \cite{robertson2024detectlensingrotation, carronDetectingRotationLensing2025c}. If a cosmic rotation happens at late-times, this might give some contamination.  The inclusion of CMB temperature could help resolve some degeneracies between these CMB distorting fields, as the curl estimator is sensitive to shear information from temperature.

Temperature can also make robust CMB lensing power spectrum analyses at very small scales (if foregrounds are under control). Recent work has shown that cosmic rotation affects the QE CMB lensing potential reconstructed power spectrum with a $N^{\hat{\phi}\alpha}_1$ bias at small scales \cite{caiImpactAnisotropicBirefringence2023b, caiEfficientEstimationRotationinduced2024a}. As de-rotation helped mitigating the bias induced by cosmic birefringence in the CMB lensing curl mode, it will be interesting to see how this affects the estimated gradient mode $\hat{\phi}$ power spectrum, particularly when including temperature data. 

To further improve our reconstruction in realistic cases, we will need to develop an end-to-end pipeline from simulations to cosmological parameters, include masking, successfully demonstrated in iterative CMB lensing \cite{legrandRobustEfficientCMB2023}, instrumental systematics that can mimic cosmological signals \citep{huBenchmarkParametersCMB2002,yadavPrimordialBmodeDiagnostics2009a, Mirmelstein_2021, collaborationBICEPKeckXVII2023}, and non-Gaussian priors \cite{darwish2024nongaussiandeflectionsiterativeoptimal}.

While here we explored in detail cosmic-birefringence, it will be interesting to consider patchy reconstruction in cross-correlation studies (e.g. with the reconstructed CMB lensing map itself, with a potential of $\mathcal{O}(10) \sigma$ level detections for CMB-S4 \cite{fengSearchingPatchyReionization2018, bianchiniInferenceGravitationalLensing2023}).

The joint-reconstruction fields affecting $B$-modes can be used to develop a more sensitive probe to primordial gravitational waves, by producing a $B$-mode template to reduce distortions affecting tensor-to-scalar ratio $r$ searches. For example, patchy re-ionization might be problematic for surveys trying to achieve $r \sim \mathcal{O}(10^{-4})$ \cite{Mukherjee_2019}. On the other hand, curl-rotation affects at the level of $\mathcal{O}(10^{-5})$ \cite{Lewis_2017}. Hence, a systematic study quantifying potential improvements in constraints on the tensor-to-scalar ratio $r$ and its uncertainty $\sigma(r)$ could be relevant for futuristic experiments like CMB-S4 pushing toward their primordial gravitational wave targets \cite{hirataAnalyzingWeakLensing2003, hirataReconstructionLensingCosmic2003, collaborationCMBS4ForecastingConstraints2022, Belkner:2023duz}.

Looking ahead, joint analysis of CMB distortions will become increasingly important for controlling systematic effects in cross-correlation studies with external tracers. For the near-future, as a first application, it will be interesting to apply standard delensing techniques to existing datasets such as SO and SPT-3G.

\begin{acknowledgments}

We are really grateful to  Julien Carron for insightful discussions, and detailed comments that helped improve the manuscript. Toshiya Namikawa, in particular for cosmic birefringence results, and best practices in CMB analyses. Sebastian Belkner for comments, and discussions about coding, gradients and priors. Federico Bianchini, in particular for sharing their code to calculate $C_L^{\tau\tau},C_L^{\tau\phi}$ in the halo model framework. We thank Lawrence Dam and Louis Legrand for providing comments to the manuscript, and Antony Lewis for feedback. OD acknowledges support from a SNSF Eccellenza
Professorial Fellowship (No. 186879). Computations were carried on the Piz-Daint CSCS cluster of the Swiss National Supercomputing Center, and the Bamboo cluster of University of Geneva.

\end{acknowledgments}

\appendix

\section{Noise theoretical calculations\label{sec:appendixnoise}}

In this section we remind how we get the theory reconstruction noise $N^0$, and the noise bias $N^1$ for the QE and iterative estimators\cite{legrandLensingPowerSpectrum2022}, as well as the residual power for cosmic rotation. For concreteness, we focus on the $EB$ estimator only.

\paragraph{Noise power} The $N^0$ bias is just the inverse of the response $R$. In the flat-sky approximation, the Gaussian reconstruction noise for the reconstructed mode $\widehat{x}(\vec{L})$ is  

\begin{multline}
  N^{0,\widehat{x}}_{L} = \\ = \left(\int_{\vec{l}_1} \frac{1}{2} \frac{1}{C_{l_1}^{EE,\mathrm{tot}}}\frac{1}{C_{l_2}^{BB,\mathrm{tot}}}f^{EB,x}(\vec{l}_1, \vec{l}_2)f^{EB,x}(\vec{l}_1, \vec{l}_2)\right)^{-1}\ ,
\end{multline}
where $\int_{\vec{l}_1} = \int \frac{d^2\vec{l}_1}{(2\pi)^2}$, $\vec{l}_2=\vec{L}-\vec{l}_1$, $f^{EB,x}_{\vec{l}_1,\vec{l}_2}=\frac{\delta \mathrm{Cov}_{\vec{l}_1,\vec{l}_2}}{\delta x_{\vec{l}_1+\vec{l}_2}}$ \cite{huMassReconstructionCMB2002, hirataReconstructionLensingCosmic2003}. Here, we get $C$ by using partially lensed (generally “operated”) spectra, by lensing (“operating”) the primordial spectra with the residual signal $C_{L}^{xx,\mathrm{itr}}=\big(1-(\rho^{\mathrm{itr}}_L)^2\big)C_L^{xx}$, the  power of the residual anisotropy, where $\rho^{\mathrm{itr}}_L$ is the cross-correlation coefficient of the reconstruction with the input at iteration $\mathrm{itr}$. This means, that for the QE we just use the lensed spectra.\footnote{An example of code is presented here \url{https://github.com/NextGenCMB/delensalot/blob/main/delensalot/biases/iterbiasesN0N1.py}.}
Similarly, we can get an $N_1$ estimate based on iterations. In the flat-sky, we write \citep{Kesden_2003, planckcollaborationPlanck2018Results2020}:
\begin{multline}
N_L^{1,\widehat{s}x}[\mathrm{itr}]= \big(R^{\widehat{s}s,\mathrm{itr}}_L R^{\widehat{s}s,\mathrm{itr}}_L\big)^{-1}\int_{\vec{l}_1\vec{l}_2} f^{EB,s}(\vec{l}_1,\vec{l}_2)f^{EB,s}(\vec{l}_3,\vec{l}_4) \\
\frac{1}{C_{l_1}^{EE,\mathrm{tot}}}\frac{1}{C_{l_2}^{EE,\mathrm{tot}}}\frac{1}{C_{l_3}^{BB,\mathrm{tot}}}\frac{1}{C_{l_4}^{BB,\mathrm{tot}}}\times 
\\\Big( C_{\vec{l}_1+\vec{l}_3}^{xx,\mathrm{itr}}f^{EB,x}(\vec{l}_1,\vec{l}_3)f^{EB,x}(\vec{l}_2,\vec{l}_4)+ \\+C_{\vec{l}_1+\vec{l}_4}^{xx,\mathrm{itr}}f^{EB,x}(\vec{l}_1,\vec{l}_4)f^{EB,x}(\vec{l}_2,\vec{l}_3)\Big)\ ,
\end{multline}
where $\vec{l}_1+\vec{l}_2=\vec{L}=-(\vec{l}_3+\vec{l}_4)$.

\paragraph{Residual rotation power} Finally, we show how to calculate the residual power spectrum $C_L^{\alpha_{\mathrm{res}}\alpha_{\mathrm{res}}}$ used to estimate the cross-correlation coefficients of $\Omega[\alpha]$ in Figure \ref{fig:theoryxcorr}. This is calculated as the difference between a normalized quadratic response, and a noise term 
\begin{equation}
 C_L^{\alpha_{\mathrm{res}}\alpha_{\mathrm{res}}} = (R^{\widehat{\alpha}\alpha})^{-2}|g^{\alpha}_{\mathrm{QD}}|^2-(R^{\widehat{\alpha}\alpha})^{-1}\ ,  \label{eq:residualpower}
\end{equation}
where $g^{\alpha}_{\mathrm{QD}}$ is the quadratic part estimating cosmic birefringence, and $(R^{\widehat{\alpha}\alpha})^{-1} = : N^{0,\widehat{\alpha}}$ is the noise. At iteration 0, we get that $C_L^{\alpha_{\mathrm{res}}\alpha_{\mathrm{res}}}=C_L^{\alpha\alpha}+N^{0,\widehat{\alpha}}-N^{0,\widehat{\alpha}}=C_L^{\alpha\alpha}$, and this is the same as when delensing only.

When derotating, we need to get $g^{\alpha}_{\mathrm{QD}}$ as a result of residual rotation in the CMB maps. In Appendix \ref{sec:appendixmf} we show that the MAP mean-field term for cosmic birefringence is small, so we ignore it. Therefore, writing $g_{\mathrm{tot}}\approx g_{\mathrm{QD}}^{\alpha}+g_{\mathrm{PR}}^{\alpha}=0$ we get $g_{\mathrm{QD}}^{\alpha} = -g_{\mathrm{PR}}^{\alpha}$.\footnote{We also check with simulations that we get a good match in the power of the prior term and the quadratic part.} The prior term is calculated as 
\begin{equation}
g_{\mathrm{PR}}^{\alpha}=-\frac{\widehat{\alpha}^{\mathrm{MAP,WF}}}{C_L^{\alpha\alpha}} = -W_L^{\alpha}\frac{\widehat{\alpha}^{\mathrm{MAP}}}{C_L^{\alpha\alpha}},
\end{equation} 
where $\widehat{\alpha}^{\mathrm{MAP}}$ is our MAP solution at iteration $\mathrm{itr}-1$, that we take to be similar as the final iteration $\mathrm{itr}$ near convergence. The Wiener filter is calculated as $W_L^{\alpha} = C_L^{\alpha\alpha}/(C_L^{\alpha\alpha}+N^{0,\widehat{\alpha}})$. This means that our residual power \ref{eq:residualpower} is 
\begin{multline}
    C_L^{\alpha_{\mathrm{res}}\alpha_{\mathrm{res}}} = \\ = \left(\frac{N^{0,\widehat{\alpha}}}{C_L^{\alpha\alpha}}\right)^2\left(\frac{C_L^{\alpha\alpha}}{C_L^{\alpha\alpha}+N^{0,\widehat{\alpha}}}\right)^2(C_L^{\alpha\alpha}+N^{0,\widehat{\alpha}})-N^{0,\widehat{\alpha}} = \\
= -\frac{N^{0,\widehat{\alpha}}C_L^{\alpha\alpha}}{C_L^{\alpha\alpha}+N^{0,\widehat{\alpha}}}=-N^{0,\widehat{\alpha}}W_L^{\alpha}
\end{multline}

\section{Mean-fields\label{sec:appendixmf}}

In this section, we examine the mean-fields obtained through our iterative MAP estimator approach. Our goal is to gauge their magnitude with respect to the signals of interest. To efficiently estimate the mean-field MAP term, we adopt the methodology proposed in Appendix B of \cite{Carron_2017}, which involves averaging Equation \ref{eq:qdpart_explicit} applied to maps of unit variance.

We calculate the mean-field arising from a single $\widehat{\xi}^{\mathrm{MAP}}$ reconstruction on the full-sky. For this section only, we generate forty-five independent maps, obtaining a set $g^{\xi}_{\mathrm{MF},i}$ where $i$ is one of the forty-five maps. While a higher number of maps leads to better results, we have chosen to use five for computational ease, as for the goals of this work we do not need much precision. 

\begin{figure}[!h]
    \centering  \includegraphics[width=0.9\linewidth]{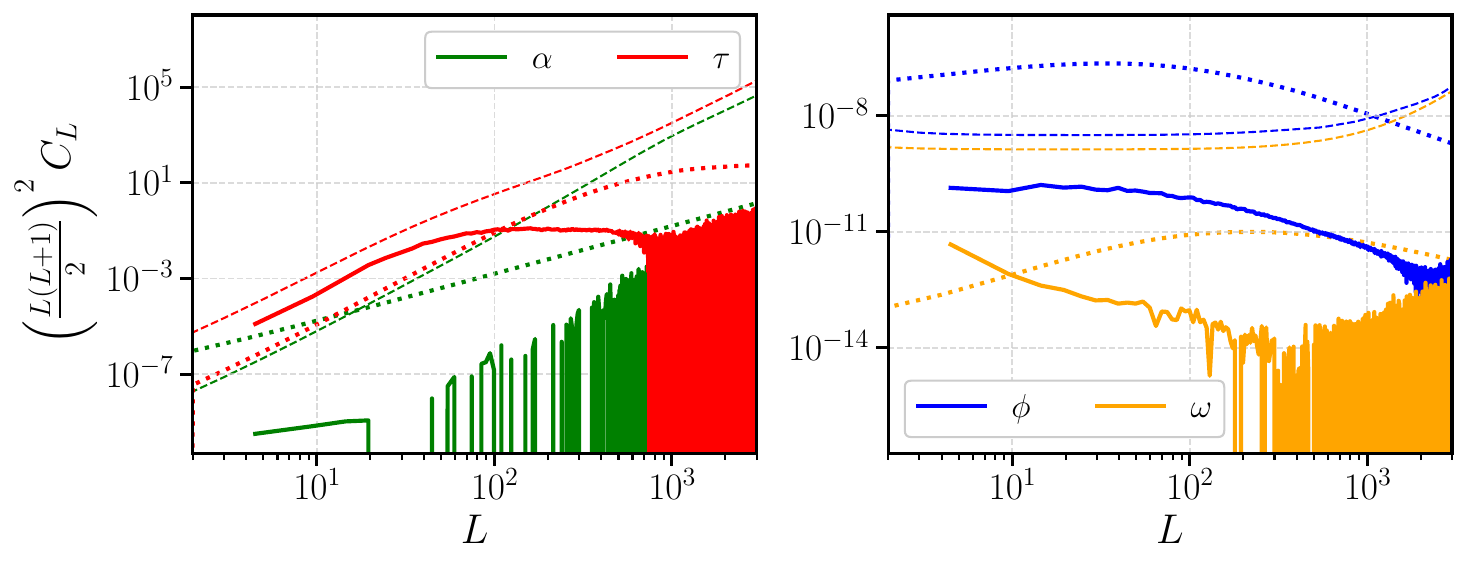}
    \caption{Mean-fields estimated at iteration $15$ and normalized with the iterated $N_0$. In green, red, blue, and orange we show the theory signal (dotted), $N_0$ noise (dashed), and mean-field power $C_L^{g^{\xi}_{\mathrm{MF}}}$, for the cosmic birefringence, patchy tau, CMB lensing gradient and curl, respectively.}
    \label{fig:meanfields}
\end{figure}

To minimize noise and obtain a clean estimate of the mean-field power spectrum, we cross-correlate results from two disjoint sets of maps:
\begin{equation}
    C_L^{g^{\xi}_{\mathrm{MF}}} = \langle g^{\xi}_{\mathrm{MF},\mathcal{A}} g^{\xi}_{\mathrm{MF}, \mathcal{B}} \rangle\ ,
\end{equation}
where $g^{\xi}_{\mathrm{MF},\mathcal{S}}=\langle g^{\xi}_{\mathrm{MF},i}\rangle_{i \in \mathcal{S}}$, $\mathcal{S}\in \{\mathcal{A},\mathcal{B}\}$, where $\mathcal{A},\mathcal{B}$ are two disjoint sets of indices. We only use one $\widehat{x}_i^{\mathrm{MAP}}$. We show results in Figure \ref{fig:meanfields}. Interestingly, $\widehat{\tau}$ and $\widehat{\Omega}$ have mean-fields comparable to their signals, on large scales, though our method is approximate. We defer to the future studying this better.

\bibliographystyle{prsty-url.bst}
\bibliography{main,joint}
\end{document}